\documentclass[preprint,12pt, authoryear]{elsarticle}

\usepackage{amssymb}
\usepackage{tabularx}

\usepackage{amsmath}
\usepackage{bm}

\usepackage{graphicx}
\usepackage{subfig}

\usepackage{array}
\newcolumntype{P}[1]{>{\centering\arraybackslash}p{#1}}
\newcolumntype{M}[1]{>{\centering\arraybackslash}m{#1}}
\usepackage{multirow}
\usepackage{booktabs,makecell}
\usepackage{float}
\usepackage[labelfont=bf]{caption}

\usepackage[colorlinks=true,allcolors=blue]{hyperref}
\usepackage[nameinlink,noabbrev,capitalize]{cleveref}

\usepackage{nomencl}
\makenomenclature
\usepackage{etoolbox}
\renewcommand\nomgroup[1]{%
  \item[\it
  \ifstrequal{#1}{S}{Symbols}{%
  \ifstrequal{#1}{G}{Greek symbols}{%
  \ifstrequal{#1}{I}{Subscripts}{%
  \ifstrequal{#1}{A}{Abbreviations}{}}}}%
]}
\usepackage{longtable}
\journal{Journal}

\begin{document}

\begin{frontmatter}

\title{Experimental and LBM analysis of medium-Reynolds number fluid flow around NACA0012 airfoil}

\author[inst1]{Andro Rak}
\ead{arak@riteh.hr}

\author[inst1,inst2]{Luka Grbčić}
\ead{luka.grbcic@riteh.hr}

\author[inst1,inst2]{Ante Sikirica}
\ead{ante.sikirica@uniri.hr}

\author[inst1,inst2]{Lado Kranjčević\texorpdfstring{\corref{cor1}}{}}
\ead{lado.kranjcevic@riteh.hr}

\cortext[cor1]{Corresponding author}

\affiliation[inst1]{organization={Faculty of Engineering, University of Rijeka},
            addressline={Vukovarska 58}, 
            city={Rijeka},
            postcode={51000 Rijeka}, 
            country={Croatia}}

\affiliation[inst2]{organization={Center for Advanced Computing and Modelling, University of Rijeka},
            addressline={Radmile Matejčić 2}, 
            city={Rijeka},
            postcode={51000 Rijeka}, 
            country={Croatia}}

\begin{abstract}

\textbf{Purpose} - 
Cornerstone of this research is the examination of fluid flow around NACA0012 airfoil, with the aim of the numerical validation between the experimental results in the wind tunnel and the Lattice Boltzmann Method (LBM) analysis, for the medium Reynolds number (Re = 191000). The LBM-Large Eddy Simulation (LES) method described in this paper opens up opportunities for faster computational fluid dynamics (CFD) analysis, because of the LBM scalability on high performance computing architectures, more specifically general purpose graphics processing units (GPGPUs), pertaining at the same time the high resolution LES approach.

\textbf{Study design/methodology/approach} - 
Process starts with data collection in open-circuit wind tunnel experiment. Furthermore, the pressure coefficient, as a comparative variable, has been used with varying angle of attack (2°, 4°, 6° and 8°) for both experiment and LBM analysis. To numerically reproduce the experimental results, the LBM coupled with the LES turbulence model, the generalized wall function (GWF) and the cumulant collision operator with D3Q27 velocity set has been employed. Also, a mesh independence study has been provided to ensure result congruence.

\textbf{Findings} - 
The proposed LBM methodology is capable of highly accurate predictions when compared to experimental data. Besides, the special significance of this work is the possibility of experimental and CFD comparison for the same domain dimensions.

\textbf{Originality/value} - 
Considering the quality of results, Root-Mean-Square Error (RMSE) shows good correlations both for airfoil’s upper and lower surface. More precisely, maximal RMSE for the upper surface is 0.105, while 0.089 for the lower surface, regarding all angles of attack.


\textbf{Paper type} Research paper
\end{abstract}

\end{frontmatter}

\nomenclature[S]{$Re$}{Reynolds number}
\nomenclature[S]{DdQq}{Velocity set with the number of spatial dimensions ($d$) and the number of discrete velocities ($q$)}
\nomenclature[S]{\textbf{$c_i$}}{Velocity set, m/s}
\nomenclature[S]{AoA}{Angle of attack, °}
\nomenclature[S]{$\overline{C_{p}}$}{Mean pressure coefficient}
\nomenclature[S]{$C_{p}$}{Pressure coefficient}
\nomenclature[S]{$p$}{Pressure, Pa}
\nomenclature[S]{$p_\infty$}{Wall static pressure, Pa}
\nomenclature[S]{$v_\infty$}{Inlet velocity, m/s}
\nomenclature[S]{$t$}{Time, s}
\nomenclature[S]{$x$}{Position in space, m}
\nomenclature[S]{$f_i$}{Particle population}
\nomenclature[S]{$f$}{Distribution function}
\nomenclature[S]{$p_{static}$}{Static pressure, Pa}
\nomenclature[S]{$p_{reference}$}{Reference pressure, Pa}
\nomenclature[S]{$p_{dynamic}$}{Dynamic pressure, Pa}
\nomenclature[S]{$p_{stagnation}$}{Stagnation pressure, Pa}
\nomenclature[S]{$c$}{Chord length, m}
\nomenclature[S]{$Dx$}{Cross section}
\nomenclature[S]{$n$}{Number of values in the sample}
\nomenclature[S]{$k$}{Turbulent kinetic energy, m\textsuperscript{2}/s\textsuperscript{2}}

\nomenclature[G]{$\xi_x$, $\xi_y$, $\xi_z$}{Particle velocity in x, y and z direction, m/s}
\nomenclature[G]{$\Omega(f)$}{Collision operator}
\nomenclature[G]{$F_{\beta}/{\rho}$}{Specific body force, N/kg}
\nomenclature[G]{$\delta{x}$}{Specific edge length, m}
\nomenclature[G]{$\delta{t}$}{Time step, s}
\nomenclature[G]{$\Delta t$}{Lattice time step, s}
\nomenclature[G]{$\sigma$}{Sample standard deviation}
\nomenclature[G]{$\omega$}{Specific dissipation rate, l/s}
\nomenclature[G]{$\epsilon$}{Turbulence dissipation rate, m\textsuperscript{2}/s\textsuperscript{3}}
\nomenclature[G]{$\rho_\infty$}{Air density in test section, kg/m\textsuperscript{3}}
\nomenclature[G]{$Re_\Theta$}{Critical Reynolds number}
\nomenclature[G]{$\gamma$}{Intermittency}

\nomenclature[I]{max}{Maximum}
\nomenclature[I]{min}{Minimum}
\nomenclature[I]{$\infty$}{Free stream}
\nomenclature[I]{$i$}{Iteration points in the data set}

\nomenclature[A]{LBM}{Lattice Boltzmann Method}
\nomenclature[A]{NACA}{National Advisory Committee for Aeronautic}
\nomenclature[A]{CFD}{Computational Fluid Dynamics}
\nomenclature[A]{LES}{Large Eddy Simulation}
\nomenclature[A]{GPGPU}{General Purpose Graphics Processing Unit}
\nomenclature[A]{SPH}{Smoothed Particle Hydrodynamics}
\nomenclature[A]{IB-LBM}{Immersed Boundary Lattice Boltzmann Method}
\nomenclature[A]{BGK}{Bhatnagar–Gross–Krook}
\nomenclature[A]{MRT}{Multiple-Relaxation-Time}
\nomenclature[A]{RANS}{Reynolds-Averaged Navier-Stokes}
\nomenclature[A]{UFX}{UltraFluidX}
\nomenclature[A]{GWF}{Generalized Wall Function}
\nomenclature[A]{VDAS}{Versatile Data Acquisition System}
\nomenclature[A]{LG}{Lattice Gas}
\nomenclature[A]{RMSE}{Root-Mean-Square Error}
\nomenclature[A]{GPU}{Graphics Processing Unit}
\nomenclature[A]{AF}{TecQuipment wind tunnel labeling}
\nomenclature[A]{AFA}{TecQuipment wind tunnel equipment labeling}
\nomenclature[A]{STD}{Standard deviation}
\nomenclature[A]{SRT}{Single Relaxation Time}
\nomenclature[A]{SST}{Shear Stress Transport}
\nomenclature[A]{RSM}{Reynolds Stress Model}
\nomenclature[A]{SA}{Spalart–Allmaras}
\nomenclature[A]{BCM}{Bas Cakmakcioglu Modified}
\nomenclature[A]{FVM}{Finite Volume Method}
\nomenclature[A]{FEM}{Finite Element Method}

\printnomenclature

\section{Introduction}
    \label{int}
    
Modeling physical phenomena using still novel LBM approach is of great interest for engineers and scientist from the computational efficiency point of view. Therefore, it is a challenging task to find the quickest, simplest and most precise way to validate the numerical results. Obviously, experimental analyses are possible, but they are expensive in contrast to numerical simulations. However, calibrating the numerical models with experimental data is a necessity in order to make sure that the numerical approach is accurate enough to model real physical phenomena. Nowadays, conventional CFD numerical methods based on solving the Navier Stokes equation, such as the finite volume method, finite difference method and finite element method, have been used the most for modeling fluid flow phenomena. However, because of increasing hardware improvements (mostly of GPGPUs), other methods, such as the LBM and the Smoothed Particle Hydrodynamics (SPH) have been increasing in popularity \citep{Kruger2017}.

Unlike methods based on solving the Navier Stokes equation, which directly obtain macroscopic variables such as pressure and velocity of fluid flow, LBM is a mesoscopic method based on solving the Boltzmann equation, which entails that it models the behavior of fluid particles \citep{Benzi1992}. It was first mentioned in the studies by \citet{Higuera1989} and \citet{McNamara1988}, where it was developed as a CFD method.

LBM is suitable for external aerodynamics, and in a last few years, there have been studies that describe fluid flow problems around airfoils, with different methods. \citet{Imamura2005} used a 2D LBM approach to model fluid flow around an airfoil for the Reynolds number of 500000. The study used a D2Q9 velocity set and investigated several angles of attack. Furthermore, the same physical problem was investigated using the immersed boundary LBM (IB-LBM) approach \citep{Peng2006, Wu2009}. The IB-LBM was further modified through the bounce-back method by \citet{Wang2020}. Also, it was used by \citet{Qiu2016} for compressible flow around an airfoil for the Mach numbers of 0.5 and 0.8.

Several other 2D LBM studies which analyse the fluid flow around an airfoil, for low Reynolds number and small variety of angle of attack, differ in the choice of the collision operator, and the used turbulence model. Most use the Bhatnagar–Gross–Krook (BGK) collision operator \citep{Luan2011, Pellerin2017, Fang2019}, however, \citet{Chen2012} applies the multiple-relaxation-time (MRT) collision operator. Regarding the used turbulence model for this physical problem, \citet{DiIlio2018} combined the Reynolds-Averaged Navier–Stokes (RANS) turbulence model with LBM, while \citet{Yao2017} introduced the LES turbulence model. \citet{ReyesBarraza2020} validated the case on various angles of attack for the low Reynolds number (up to 12 000), while using BGK collision operator. \citet{Zhuo2010}, used an LBM approach with an increased Reynolds number up to 1 million, with a difference in velocity set as D2Q13, instead \citet{ReyesBarraza2020} D2Q9. Furthermore, \citet{Li2012} combined the MRT collision operator with LES, while \citet{Pellerin2015} used a cascaded collision operator with a RANS turbulence model. Both studies used a Reynolds number of 500000. Finally, \citet{Hejranfar2018} analyzed the fluid flow around an airfoil for the Reynolds number of 6 million with the D2Q9 velocity set.

\citet{Leveque2018}, in contrast to others, introduces 3D solution of the problem with D3Q19 velocity set and BGK collision operator. They validate the case on just one angle of attack, for low Reynolds number (up to 42 000). \citet{Degrigny2021} used the same LBM-LES approach to investigate fluid flow around the airfoil for high Reynolds numbers. Finally, \citet{Wilhelm2018} investigated the same problem with an LBM-RANS coupled model, but for a different airfoil.

The common thing between the previously mentioned high Reynolds flow studies is that all of them incorporated the D3Q19 velocity set. With the increase of computational resources, especially GPGPUs, it is possible to use the more computationally intensive D3Q27 velocity set, and instead of the most common BGK collision operator, investigate a more complex variant (such as MRT or cumulant collision operator) on the fluid flow around an airfoil problem.

In this study, we present and validate a complete 3D LBM modelling framework for the analysis of fluid flow around the NACA0012 airfoil. Firstly, an experimental analysis of the NACA0012 airfoil is conducted in an open-circuit wind tunnel in order to validate the numerical approach. The angle of attack (AoA) is varied (2°, 4°, 6° and 8°) and a medium Reynolds number is set at the inlet (Re = 191000). Furthermore, using Altair’s UltraFluidX (UFX) solver, the LBM-LES model with a D3Q27 velocity set and the high-fidelity Cumulant-based collision operator is applied to numerically investigate the fluid flow around the airfoil. Also, the GWF is used for wall treatment. The numerical approach is compared with the obtained experimental data, and a mesh independence analysis is presented. The experimental setup is elaborated in \cref{exp}, while the numerical setup with an LBM theoretical background is covered in \cref{num_model}, the comparison of numerical and experimental data is revealed in \cref{res}.

\section{Materials and Methods}


\subsection{Experimental setup}
    \label{exp}
The experiment was conducted using Techquipment’s ISO 9001 certified open-circuit suction AF1300 subsonic wind tunnel with a test section of 305mm by 305mm and 600mm long. The tunnel is constructed to absorb air from the atmosphere through an aerodynamically designed, conical contractor. It speeds up air linearly due to a difference in cross sections. A honeycomb-shaped section at the inlet in order to increase the uniformity of air flow. An axial fan that retracts air back into the atmosphere is positioned after the diffuser. The wind tunnel has a length of 3700mm, a width of 1065mm and a height of 1900mm.

\begin{figure}[H]
      \centering
      \includegraphics[width=0.78\columnwidth]{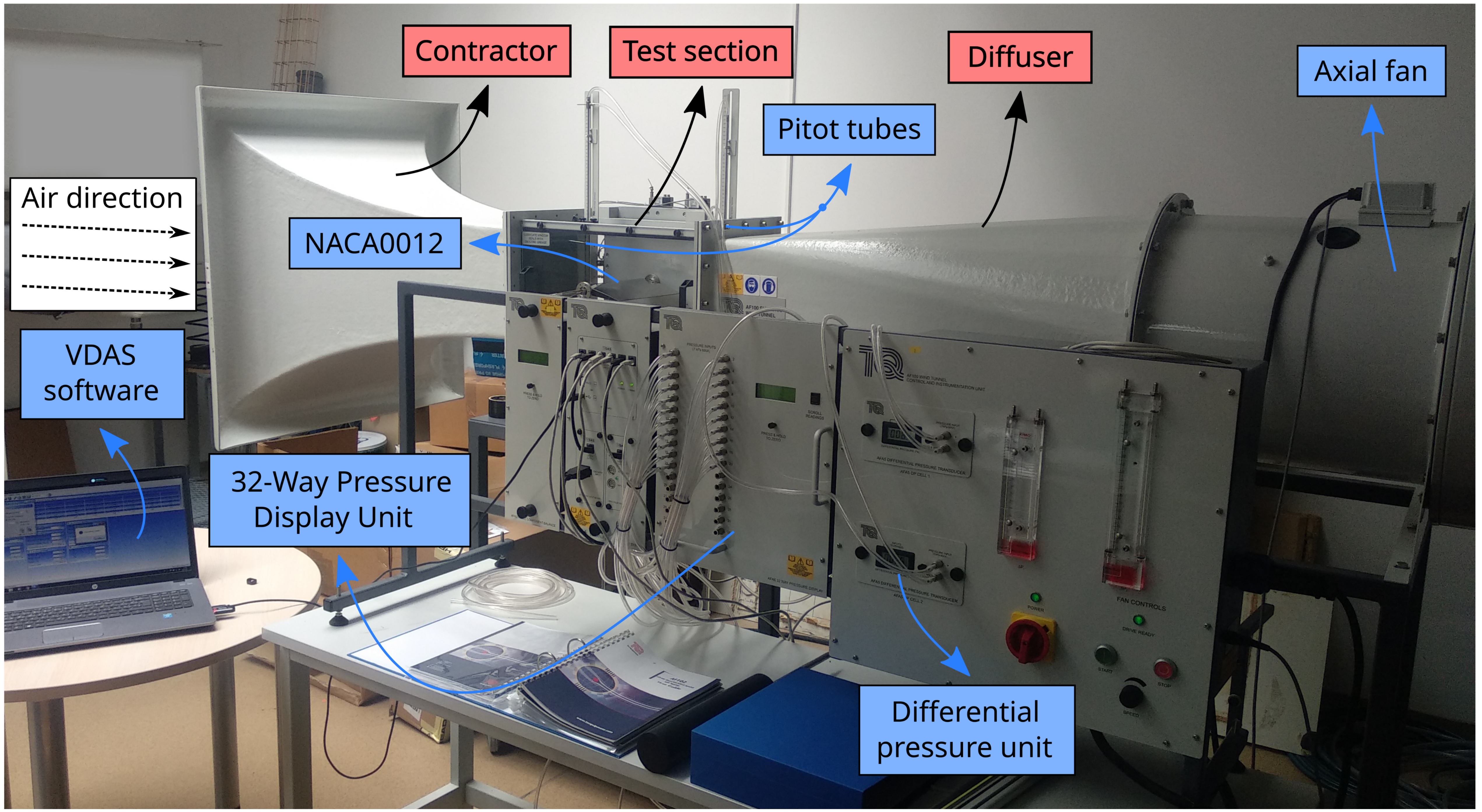}
      \caption{Subsonic open-circuit wind tunnel with three main parts from where the air goes through are highlighted red, while other sensors, control panel, software, airfoil and fan are displayed with blue. \label{WT}}
\end{figure}

Experimental data is obtained with several ISO 9001 certified Techquipment’s sensors located within the test section, of which wind tunnel components are displayed in \cref{WT}. Two Pitot-Static tubes are connected to the AFA5 differential pressure unit, with the range of ± 7 kPa. Also, 20 static pressure tappings are connected to the manifold, from where they go to the AFA6 32-Way Pressure Display Unit, with the range of ± 7 kPa, with the same merging process as the wall static pressure tap. Finally, a protractor and a model holder for angle adjustment are located at the back of the test section.

Experimental procedure starts with setting up the Pitot-static probe at the upstream position, to 30 mm from the top of the test section. Firstly, an optimal position of the Pitot-static tube is calibrated in order to minimize boundary layer effects in a stabilized fluid flow. After the calibration procedure, the NACA0012 airfoil has been positioned inside the test section and analysed with a Reynolds number of 191000. The NACA0012 airfoil dimensions are 150 mm chord length and a 300 mm span length. Also, the airfoil contains 20 static pressure tappings along the chord, which are geometrically equally distributed on the upper and lower surface. Those geometrical distributions are presented in \cref{airfoil}, with the exact distance from the leading edge for every node.

\renewcommand*\thetable{\Roman{table}}
\begin{table}[H]
\renewcommand{\arraystretch}{1.2}
\caption{Tapping position for pressure measurement on NACA0012 airfoil.\label{airfoil}}
\newcolumntype{C}{>{\centering\arraybackslash}X}
\begin{tabularx}{\textwidth}{CCCC}
\toprule
\textbf{Upper surface tapping}	& \textbf{Distance from leading edge [mm]}	& \textbf{Lower surface tapping} & \textbf{Distance from leading edge [mm]}\\
\midrule
1		& 0.76			& 2        & 1.52\\
3		& 3.81			& 4        & 7.62\\
5		& 11.43			& 6        & 15.24\\
7		& 19.05			& 8        & 22.86\\
9		& 38.00			& 10       & 41.15\\
11		& 62.00			& 12       & 59.44\\
13		& 80.77			& 14       & 77.73\\
15		& 101.35		& 16       & 96.02\\
17		& 121.92		& 18       & 114.30\\
19		& 137.16		& 20       & 129.54\\
\bottomrule
\end{tabularx}
\end{table}

In order to obtain proper experimental results, a few steps are necessary. Firstly, the trailing edge of the airfoil has been positioned at the same height as the centre line of the model holder. Secondly, the tube connections are checked and after the fluid flow stabilizes, sensors start taking experimental values. TecQupiment’s Versatile Data Acquisition System (VDAS) software is used to record and export experimental data. The relevant experimental values are recorded every 0.5 seconds for a total of 300 seconds. All recorded values are time averaged, and in order to carry out a quantitative analysis, the pressure coefficient $C_{p}$ (defined in \cref{Cp}) has been defined as a benchmark for further comparison and validation. $C_{p}$ is a non-dimensional variable that defines the ratio of relative pressure difference over the kinetic pressure of a fluid.

\begin{equation}
    C_p = \frac{p - p_\infty}{\frac{1}{2} \rho_\infty v^{2}_\infty}
    \label{Cp}
\end{equation}

The symbol $\infty$ means that free stream pressure, velocity and density moves away from the airfoil, i.e. the air passing through the test section. Hence, $p_\infty$ becomes wall static pressure, while $p$ belongs to the pressure at the tapping points on the airfoil profile. $v_\infty$ specifies velocity at the inlet of the test section, while $\rho_\infty$ defines medium density inside the test section. The wind tunnel test section operating conditions are: atmospheric temperature of 27°C and atmospheric pressure of 1015.2 mbar. Furthermore, the ambient air density is 1.18 kg/m\textsuperscript{3} and the inlet air velocity is 20 m/s. The experiment has been carried out for four different angles of attack, 2°, 4°, 6° and 8°.


\subsection{Numerical model}
    \label{num_model}
The Altair UFX LBM solver was used to investigate the accuracy of the LBM method on the fluid flow around the NACA0012 airfoil. Also, the solver uses the LES turbulence model. The theoretical aspects of the LBM implementation are presented in \cref{LBM_sec}, while the detailed numerical setup is explained and presented in \cref{Num_setup}.

\subsubsection{Lattice Boltzmann method}
    \label{LBM_sec}
The Lattice Boltzmann Method was derived as a solution to the particle-based Lattice gas models, which suffered from statistical noise. The dynamics of Lattice gas (LG) are too complex to model the fluid flow in terms of macroscopic variables, hence microscopic models aren’t suitable for CFD modeling. LBM is a mesoscopic method, as it tracks the particle distribution, and not individual particles \citep{Tovbin1990, Silva2019}.

\begin{figure}[H]
      \centering
      \includegraphics[width=0.99\columnwidth]{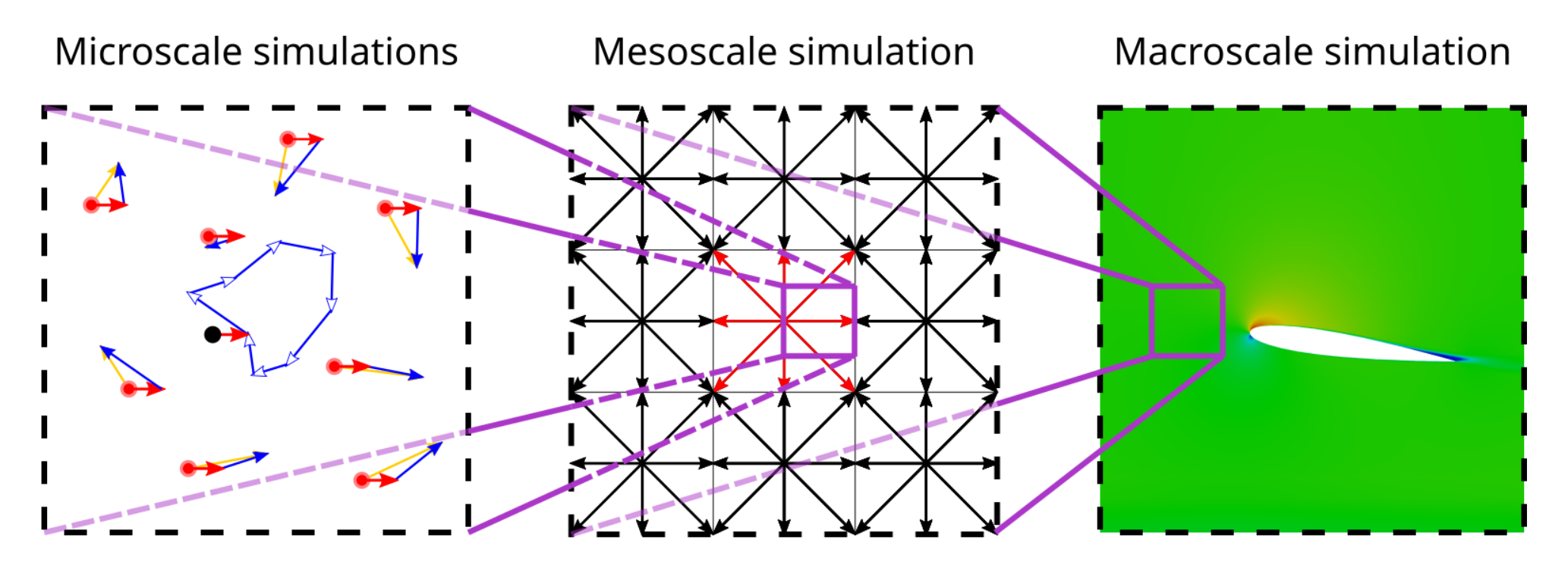}
      \caption{The mesoscale simulation (middle) is a simplification of microscale simulation (left) which allows the acquisition of macroscale simulation variables (right). \label{meso}}
\end{figure}

A microscopic description of particles entails tracking the particles on a molecular level, however, observing particles in a defined area allows for a simplification. The left part of \cref{meso} reveals that statement. Every local particle in the observed area has a velocity (yellow vector). Furthermore, when all vectors are summed up and divided by the number of particles in the constrained area, the average or flow velocity (red vector) is obtained. Each particles have an individual velocity deviation from the average velocity, which reflects chaotic motion (blue vector). These velocity deviations yield zero when summed and can be neglected (light blue vector).

The LBM method tracks the particle distribution instead of each individual particle which gives an advantage in terms of computational time and accuracy (middle \cref{meso}). LBM is based on kinetic theory, where the fundamental variable is the particle distribution function $f(x, \xi, t)$. Function represents the density of particles with the velocity vector of three components $\xi{x}$, $\xi{y}$ and $\xi{z}$ at position $x$ and time $t$. The time derivative of the particle distribution gives the following \cref{equationname}:

\begin{equation}
    \frac{df}{dt} = \left( \frac{\partial{f}}{\partial{t}} \right) \frac{dt}{dt} + \left( \frac{\partial{f}}{\partial{x_{\beta}}} \right) \frac{dx_{\beta}}{dt} + \left( \frac{\partial{f}}{\partial{\xi{_{\beta}}}} \right) \frac{d\xi{_{\beta}}}{dt}
    \label {equationname}
\end{equation}

With taking moments from the particle distribution functions, which is the link between the mesoscopic and the macroscopic view, the possibility opens up for obtaining the macroscopic pressure and velocity of the fluid. From \cref{equationname}, particle velocity should be written as $dx_{\beta}/dt=\xi_{\beta}$, while the specific body force from the Newton’s second law is $d\xi_{\beta}/dt=F_{\beta}/{\rho}$. Also, for changing the total differential with the collision operator as $\Omega(f)=df/dt$, the Boltzmann equation is obtained (\cref{Boltzmann}):

\begin{equation}
    \frac{\partial{f}}{\partial{t}} + {\xi_{\beta}} \frac{\partial{f}}{\partial{x_{\beta}}} + \frac{F_\beta}{\rho} \frac{\partial{f}}{\partial{\xi{_{\beta}}}} = \Omega(f)
    \label {Boltzmann}
\end{equation}

The Boltzmann equation can be perceived as an advection equation (hyperbolic type) where the first two terms represent the distribution function, while the third term represents forces. Also, the source term or the collision operator is on the right-hand side of the equation, and it describes the redistribution of particles due to particle collision. The numerical discretization scheme for solving the Boltzmann equation is simple for implementation and parallelization. Also, the collision operator $\Omega(f)$ depends on the local value of $f$ and not its gradients \citep{Kruger2017}.

Discrete-velocity distribution function $f_i$, better known as particle populations, is an important part of the LBM method. The major difference between particle populations $f_i$ and distribution function $f$ is discretization, i.e. argument variables of $f$ are continuous, while for $f_i$ are discrete. Hence, to get the LBM equation, the necessity is to discretize the Boltzmann equation in velocity space, physical space, and time \citep{Chen1998, Kruger2017}. Therefore, a discrete location and time are needed. The UFX solver discretizes the space into individual voxels with the specific edge length ${\delta}x$. Methodology is similar to the classical finite difference method. Time is, on the other hand, discretized into time steps, ${\delta}t$. Consequently, the distribution function is defined at the centroids of each voxel at every time step. More details about the grid refinement with a setup for the NACA0012 analysis are provided in \cref{Num_setup}.

\begin{figure}[H]
      \centering
      \includegraphics[width=0.45\columnwidth]{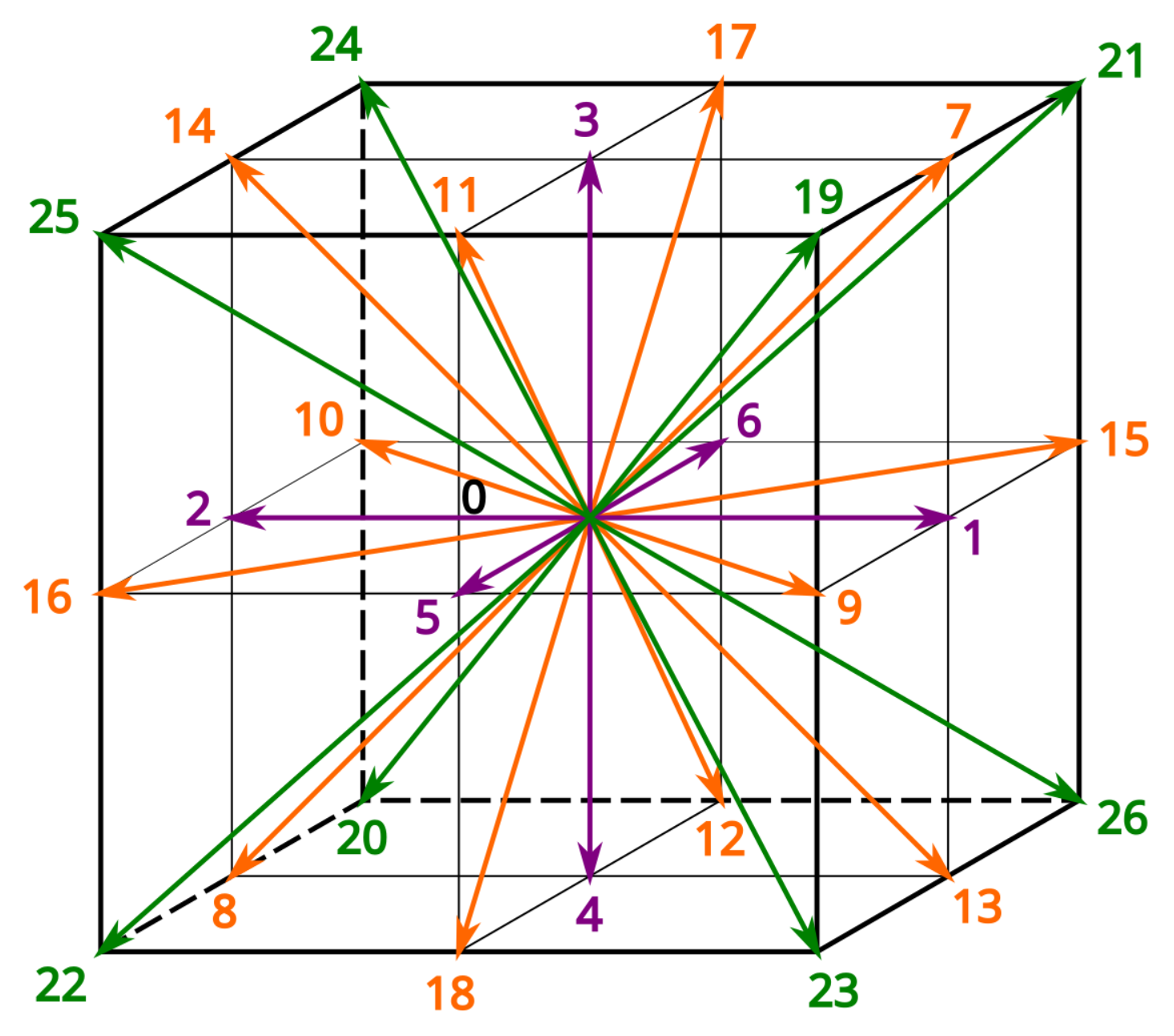}
      \caption{D3Q27 velocity set that is defined with one zero velocity at the center and 26 non-zero velocities. \label{vel_set}}
\end{figure}

The Boltzmann equation is also discretized in velocity space, which leads to determining the velocity set $\{ c_i \}$. Nevertheless, each velocity set uses the notation $Dd/Qq$, where $d$ represents the number of spatial dimensions and $q$, the number of discrete velocities. UFX uses the D3Q27 velocity set, which is illustrated in \cref{vel_set} \citep{Kruger2017, Lallemand2021, Over2021}.

The D3Q27 provides higher numerical stability (significantly less numerical errors) but requires more computing power. Previous studies have shown the robustness of the D3Q27 velocity set when it comes to modeling medium and high Reynolds number flows \citep{White2011, Suga2015, Kang2013}. Due to given information, the D3Q27 velocity set should be more accurate than other popular velocity sets, such as D2Q9, D3Q15 and D3Q19. When discretized, the Lattice Boltzmann equation is defined as (\cref{LB}):

\begin{equation}
    f_i \left( \textbf{x} + \textbf{c}_i\Delta t, t + \Delta t \right) = f_i \left( \textbf{x}, t \right) + \Omega_i(\textbf{x}, t)
    \label {LB}
\end{equation}

\begin{figure}[H]
      \centering
      \includegraphics[width=0.78\columnwidth]{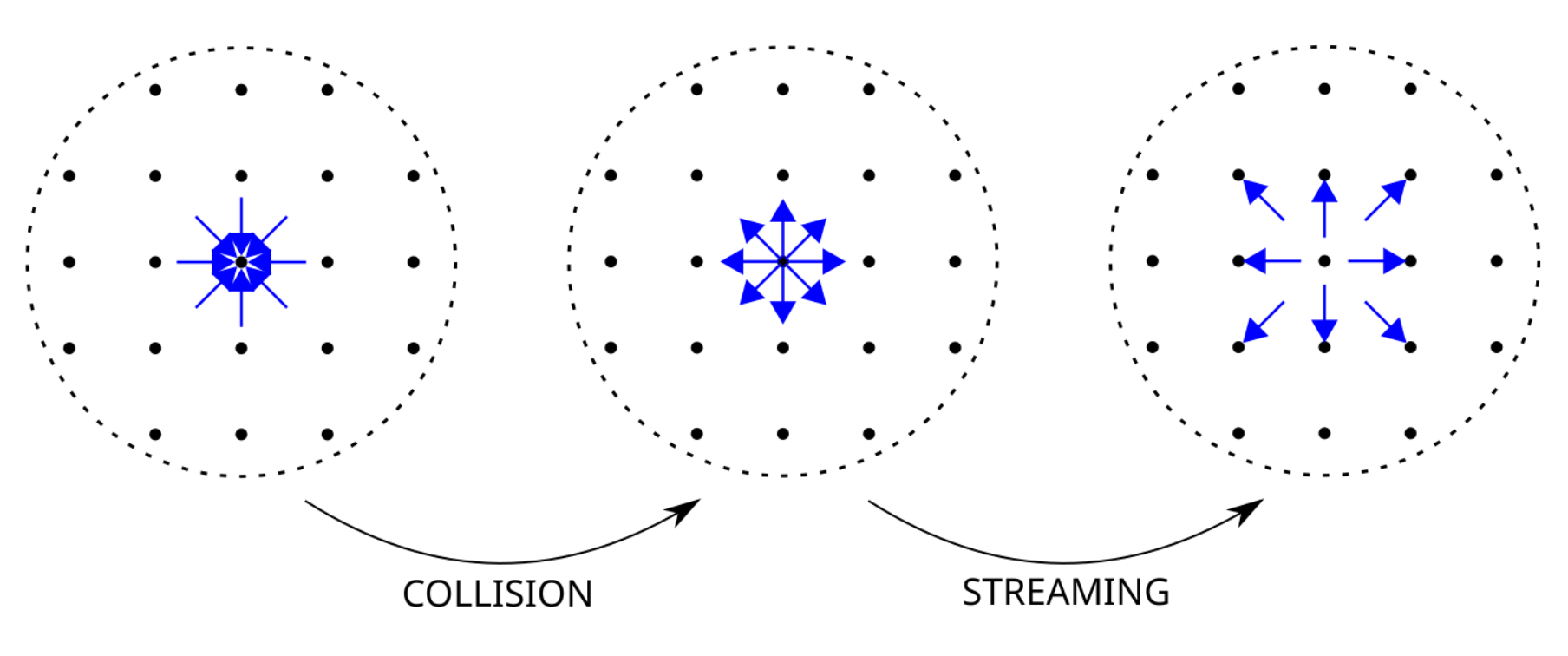}
      \caption{LBM process of collision and streaming of particles represented in 2D by the D2Q9 velocity set. \label{collision}}
\end{figure}

\cref{LB} defines the particle movement with the velocity $\textbf{c}_i$ towards the neighboring node ($\textbf{x} + \textbf{c}_i\Delta t$) in the time step ($t + \Delta t$). The collision operator $\Omega_i(\textbf{x}, t$) represents the particle collisions during each time step. More generally, \cref{LB} comprises of two parts: collision (relaxation) and streaming (propagation) of particles, both shown in \cref{collision}. Firstly, the particles collide (left part of \cref{collision}), and afterwards the particles move from the center node towards their adjacent node. After each collision, the particle distribution function is changed. Secondly, the streaming part occurs, as shown on the right part of \cref{collision}, where the particles propagate towards their neighbouring nodes.

Appearance and complexity of the two main parts depends on the collision operator $\Omega_i(\textbf{x}, t)$, that can influence the numerical properties of the LBM method \citep{Coreixas2020}. UFX uses the high-fidelity cumulant collision operator with very low numerical diffusion \citep{Kutscher2019, Geier2015, Geier2017, Geier2017a, Pasquali2020}. Moreover, the cumulant collision operator is newer, highly accurate and more complex, than e.g. the BGK operator \citep{Bhatnagar1954, Filippova1998, Tamura2011, Mendu2012, Nathen2018, Jacob2019, Hejri2020}.

UFX applies the LES turbulence model \citep{Sagaut2006, Ferziger2002, Moser2021, Bose2018} with a sub-grid scale Smagorinsky model \citep{Asmuth2021, Dong2008, Sajjadi2017, Haussmann2019}. Furthermore, UFX uses the turbulent boundary layer equation near the wall \citep{Bose2018, Chen2014}. The boundary schemes in LBM are classified in two groups: link-wise and wet-node, and both are implemented in the UFX solver. Link-wise scheme at solid boundaries and wet-node at inlet or outlet boundary. The bounce-back method \citep{Bouzidi2001} is used for the link-wise boundary scheme \citep{Aidun2010}, which means that during particle propagation, the particles reflect to their original location with reversed velocity when they collide with a solid boundary \citep{Kruger2017, Over2021}.

\subsubsection{Numerical setup}
    \label{Num_setup}
\begin{figure}[H]
      \centering
      \includegraphics[width=0.65\columnwidth]{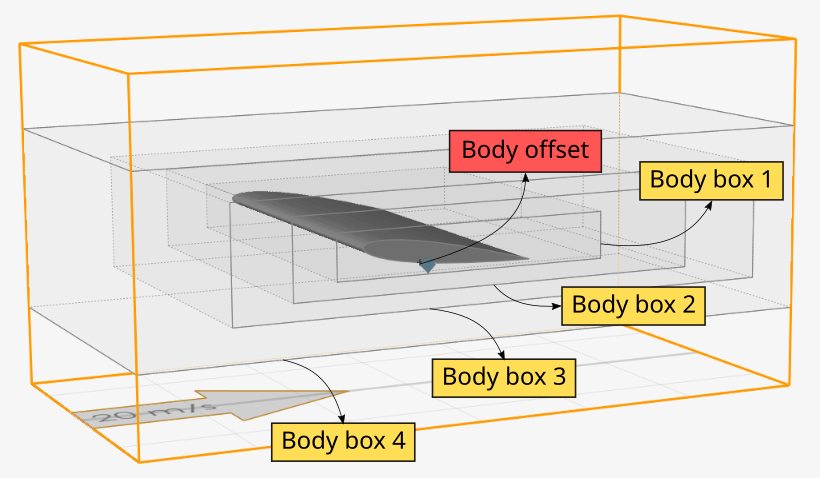}
      \caption{Position of body boxes (transparent grey orthotopes) and body offset (triangle that symbolises the mesh refinement around the object) around the airfoil and inside the wind tunnel numerical domain. \label{VWT}}
\end{figure}

The numerical domain is defined to be identical to that of the wind tunnel test section of the experiment (0.6m x 0.305m x 0.305m). The NACA0012 profile has the same chord length (0.15m) and is identically positioned inside the test section, as in the experimental analysis. Furthermore, mesh controls have been defined to accurately determine the airfoil boundary inside the domain. More precisely, four refinement level body boxes and a body offset are depicted in \cref{VWT}.

Each refinement level box defines a mesh size. To define the whole mesh, the coarsest mesh size has to be defined. Afterwards, at each following box has double refinement. A more detailed review of the refinement boxes and body offset is presented in \cref{box}.
\begin{table}[H]
\renewcommand{\arraystretch}{1.2}
\caption{Body boxes distribution in domain with their dimensions and refinement level with included body offset distance from the object. \label{box}}
\newcolumntype{C}{>{\centering\arraybackslash}X}
\begin{tabularx}{\textwidth}{CCCC}
\toprule
\textbf{Mesh control type}	& \textbf{Refinement level}	& \textbf{Type} & \textbf{Offset Distance [m]}\\
\midrule
Body Offset		& 6			& Distance      & 0.002\\
\bottomrule
\textbf{Mesh control type}	& \textbf{Refinement level}	& \textbf{Dimensions (L x W x H) [m]} & \textbf{Position (X x Y x Z) [m]}\\
\midrule
Body Box 1		& 4			& 0.24 x 0.305 x 0.04       & 0.170 x 0.0 x 0.129\\
Body Box 2		& 3			& 0.36 x 0.305 x 0.07       & 0.132 x 0.0 x 0.115\\
Body Box 3		& 2			& 0.48 x 0.305 x 0.10       & 0.080 x 0.0 x 0.100\\
Body Box 4		& 1			& 0.60 x 0.305 x 0.16       & 0.000 x 0.0 x 0.070\\
\bottomrule
\end{tabularx}
\end{table}

\begin{figure}[H]
      \centering
      \includegraphics[width=0.6\columnwidth]{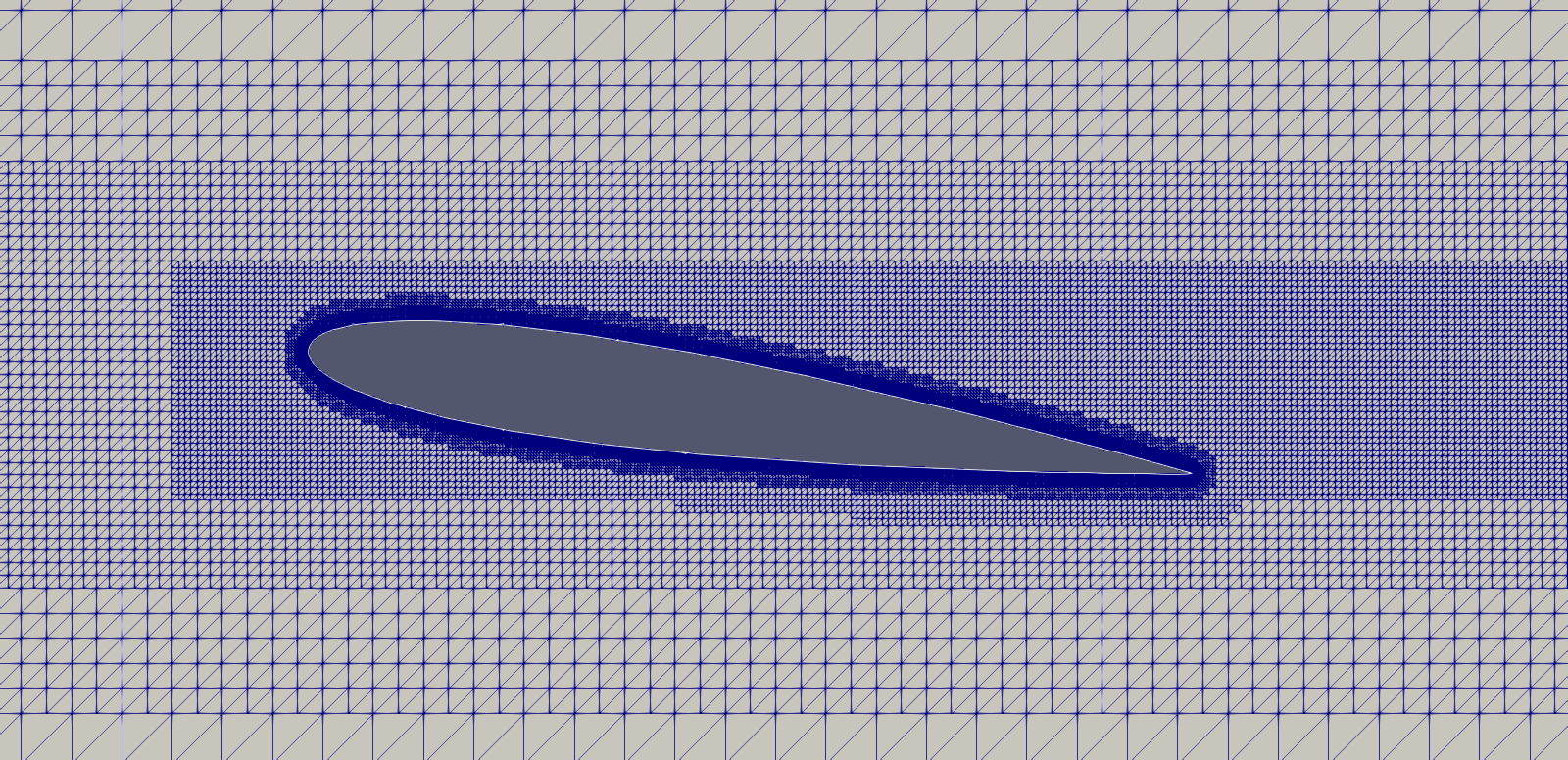}
      \caption{Mesh distribution around NACA0012 airfoil with consideration of refinement boxes and body offset.\label{box_mesh}}
\end{figure}

The mesh refinement around the airfoil can be seen in \cref{box_mesh}, where the initial mesh has been defined. The mesh is gradually refined towards the airfoil, and hence, it can be concluded that the flow resolution will be higher in a smaller sized voxel.

To complete the numerical setup, several other parameters need to be defined, the fluid is defined as air with the properties identical to those in the experimental analysis, i.e. the temperature is set to 27°C, the density 1.18 kg/m\textsuperscript{3} and velocity 20 m/s. Also, the one-way coupling with the GWF was applied near the wall \citep{Shih2003}. The slip velocity factor is set to 0.5, which numerically enforces a thinner boundary layer of the wall model. Furthermore, the number of the coarsest iterations is automatically defined regarding inflow velocity (20 m/s), Mach factor (1) and runtime (0.3 s). UFX calculates the runtime as (30*chord length)/inflow velocity, but it is extended from 0.225 s to 0.3 s to ensure numerical stability. The authors have investigated the fluid flow around the airfoil for longer simulation runtimes, however, the results remained identical. In order to get valid and stable results, only the last 10\% of the solution is considered, i.e. the last 10\% of obtained variables are time averaged. To assign this task through UFX, the following parameters have been used as \textit{average start coarsest iteration} = \textit{number of coarsest iterations} * 0.9, where the first parameter designates the starting point from where on results will be averaged.

\begin{figure}[H]
      \centering
      \includegraphics[width=0.7\columnwidth]{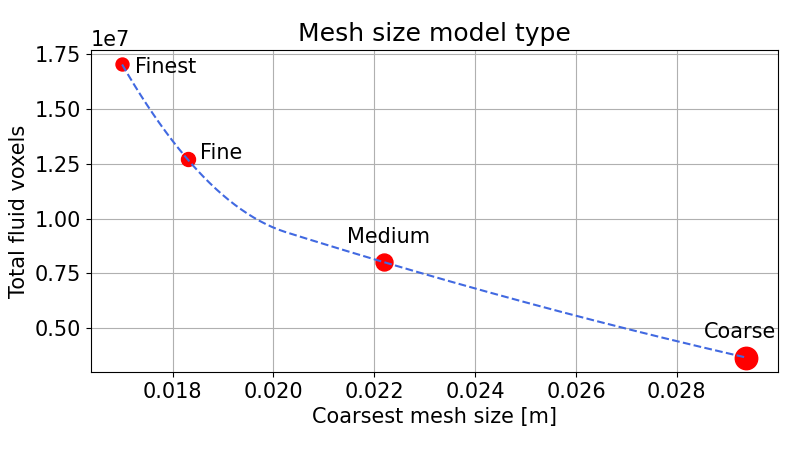}
      \caption{Voxels growth in correlation with coarsest mesh size between 4 different mesh models (Coarse, Medium, Fine and Finest), with their individual coarsest mesh size. \label{voxel}}
\end{figure}

Four different meshes (from coarsest to finest) are investigated and their details, in terms of voxel number and individual coarsest mesh size ratio, are revealed in \cref{voxel}.

It is necessary to obtain a mesh independent solution, hence the models were investigated on the four different meshes in terms of size. Results of this mesh independence analysis will be provided in \cref{mesh_ass}.

\section{Results and discussion}
    \label{res}


\subsection{Experimental results}
    \label{exp_res}
Results of the experimental measurements have been analyzed and evaluated in Python 3.9. Experimental data will be presented via the pressure coefficient. The pressure coefficient has been calculated according to \cref{Cp}, defined in \cref{exp}. Equation incorporates static pressure, stagnation pressure and reference pressure and can be further defined as (\cref{Cp_stag}):

\begin{equation}
    C_p = \frac{p_{static} - p_{reference}}{p_{stagnation} - p_{reference}} = \frac{p_{relative}}{p_{dynamic}}
    \label{Cp_stag}
\end{equation}

where $p_{reference}$ is the freestream static pressure inside the test section, while $p_{relative}$ is the divergence static pressure. Additionally, $p_{dynamic}$ is difference between stagnation and reference pressure. Use of $C_p$ enables clear representation of the relative pressure distribution. The maximum value $C_p$ can achieve is $C_p$ = 1, which means that the static pressure and stagnation pressure are equal i.e. $p_{static} = p_{stagnation}$, which is true when the fluid is at rest. If $p_{static} = p_{reference}$, then the pressure coefficient, $C_p$ = 0.

\begin{figure}[H]
      \centering
      \includegraphics[width=0.65\columnwidth]{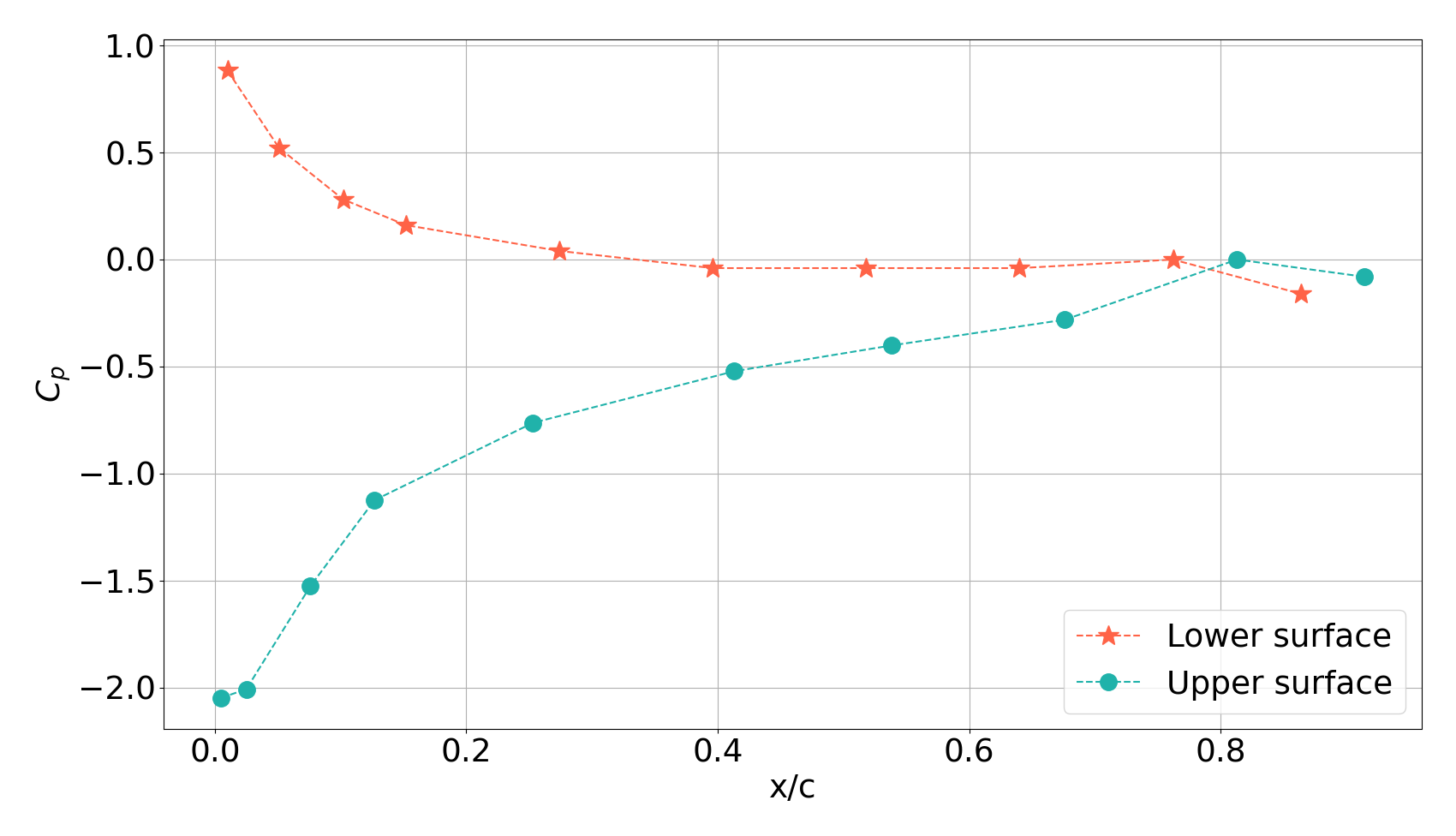}
      \caption{Measured $C_p$ at tapping points on lower and upper surface for the angle of attack of 8°.\label{Cp_8_graph}}
\end{figure}

Typically, when plotting pressure coefficient graphs, abscissa is the ratio of leading edge and chord length, x/c, of the airfoil. This approach is adopted and a reference plot is presented in \cref{Cp_8_graph}.

\cref{Cp_8_graph} demonstrates asymptotic behavior of the pressure coefficient values with regard to abscissa when approaching the trailing edge. However, there is a notable deviation at the trailing edge, near x/c = 0.8. It is important to note that deviation takes place at the part of the airfoil profile where separation occurs and a considerable turbulent region is formed. Furthermore, used open wind tunnel as well as measurement equipment contribute to the disparity. Undoubtedly, induced vibrations can have a notable impact, especially at higher speeds, since airfoil has cantilever reception. Surface roughness imperfection, as described by \citet{Ye2021}, should also be considered.

\begin{figure}[H]
    \centering
    \hfill
    \subfloat[Lower surface.]{\includegraphics[width=0.48\linewidth]{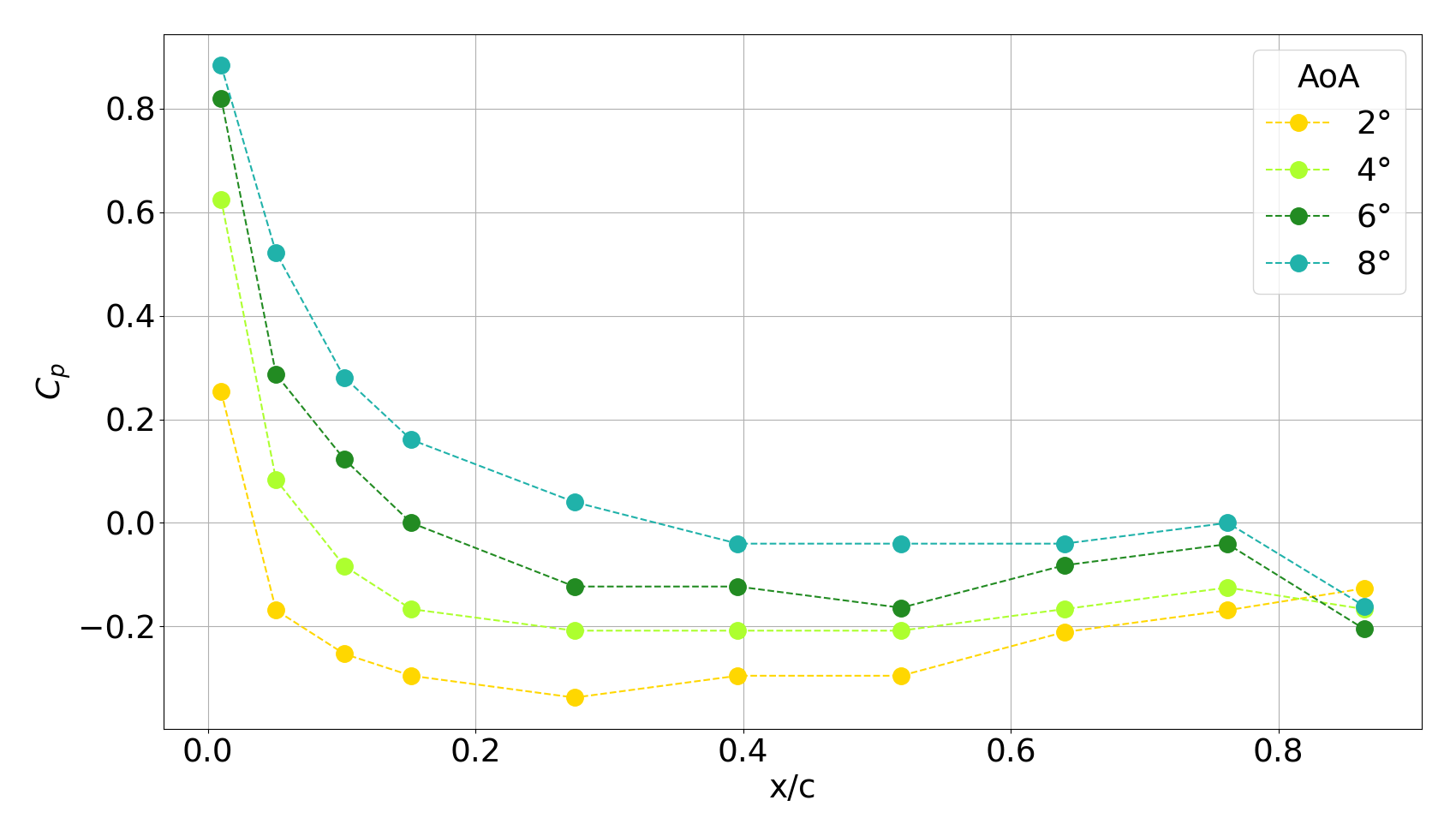}\label{Low_Cp}}
    \medskip
    \subfloat[Upper surface.]{\includegraphics[width=0.48\linewidth]{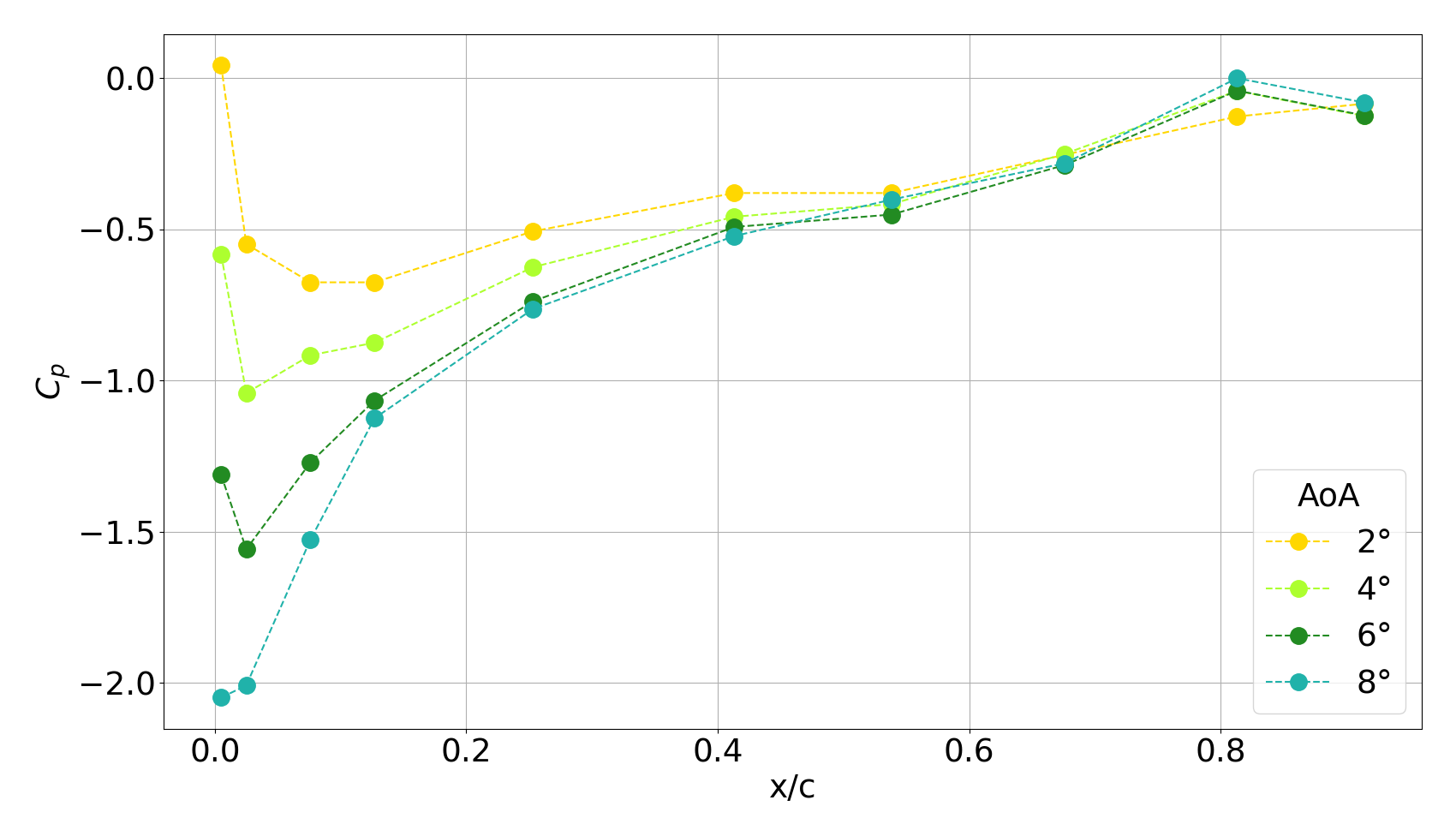}\label{Up_Cp}}
    \caption{$C_p$ distribution over airfoil for different AoA.\label{Cp_Exp}}
\end{figure}

Experimental results do not reach $C_p$ = 1 in our testing since, due to technological limitations, tap locations had to be moved downstream on both lower and upper surface; this effectively means that the initial measurement points start slightly downstream of the x/c = 0 point, where largest $C_p$ value is expected. Experimental results for $C_p$ at different AoA are shown in \cref{Cp_Exp}.

\begin{figure}[H]
  \centering
  \subfloat[AoA = 2°.]{\includegraphics[width=0.48\linewidth]{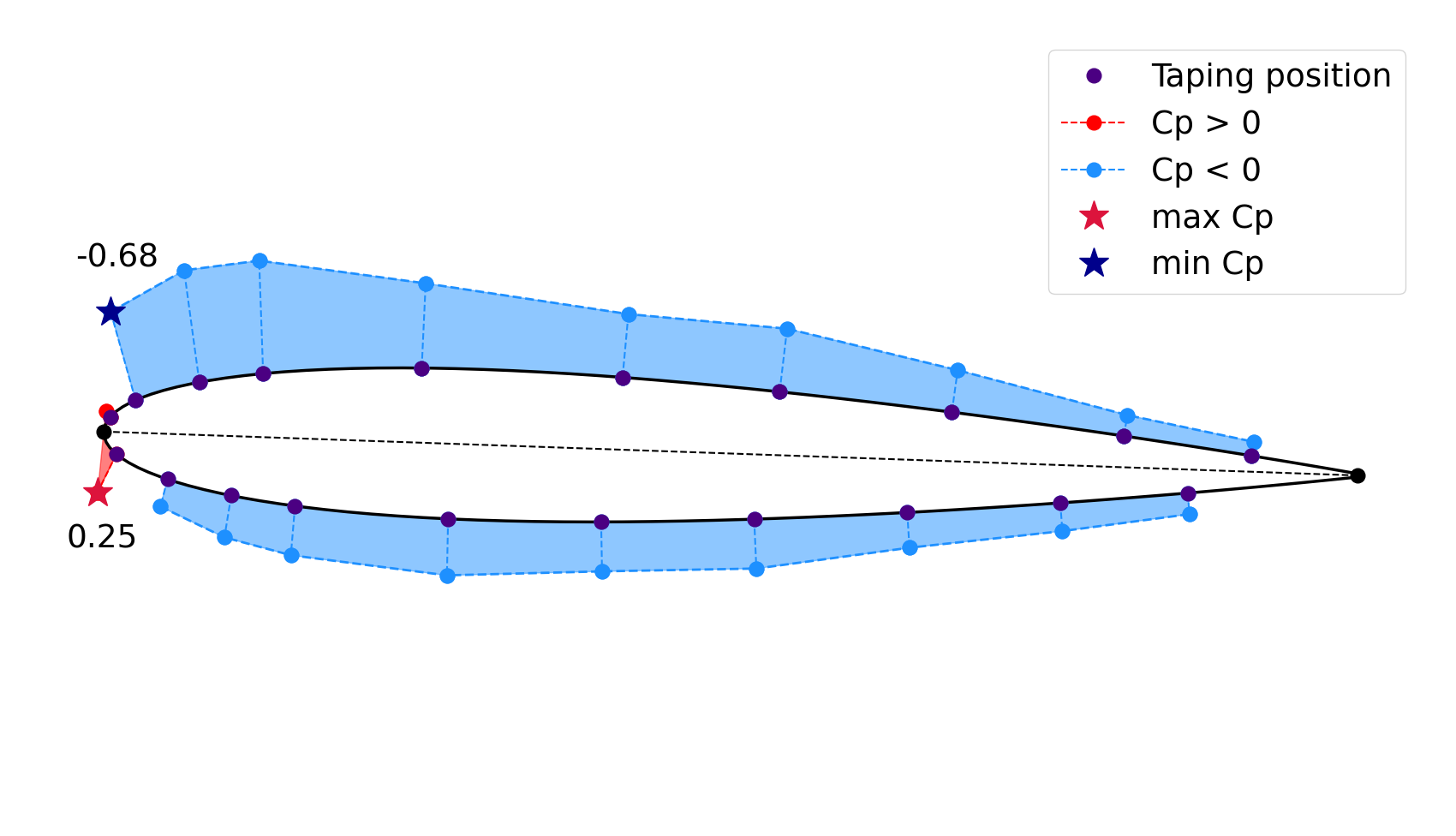}\label{Cp_dis2}}
  \medskip
  \subfloat[AoA = 4°.]{\includegraphics[width=0.48\linewidth]{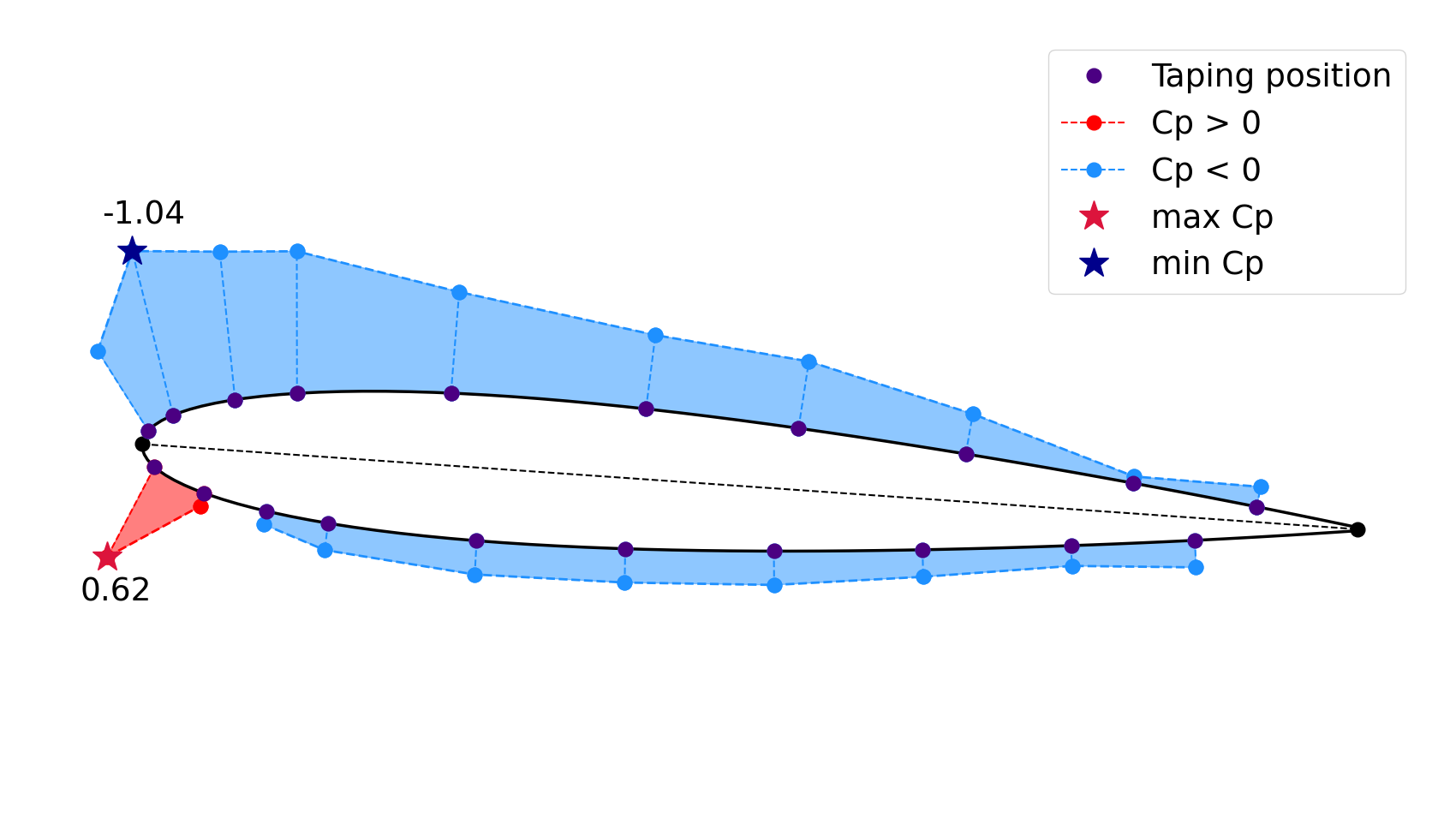}\label{Cp_dis4}}
  \hfill
  \subfloat[AoA = 6°.]{\includegraphics[width=0.48\linewidth]{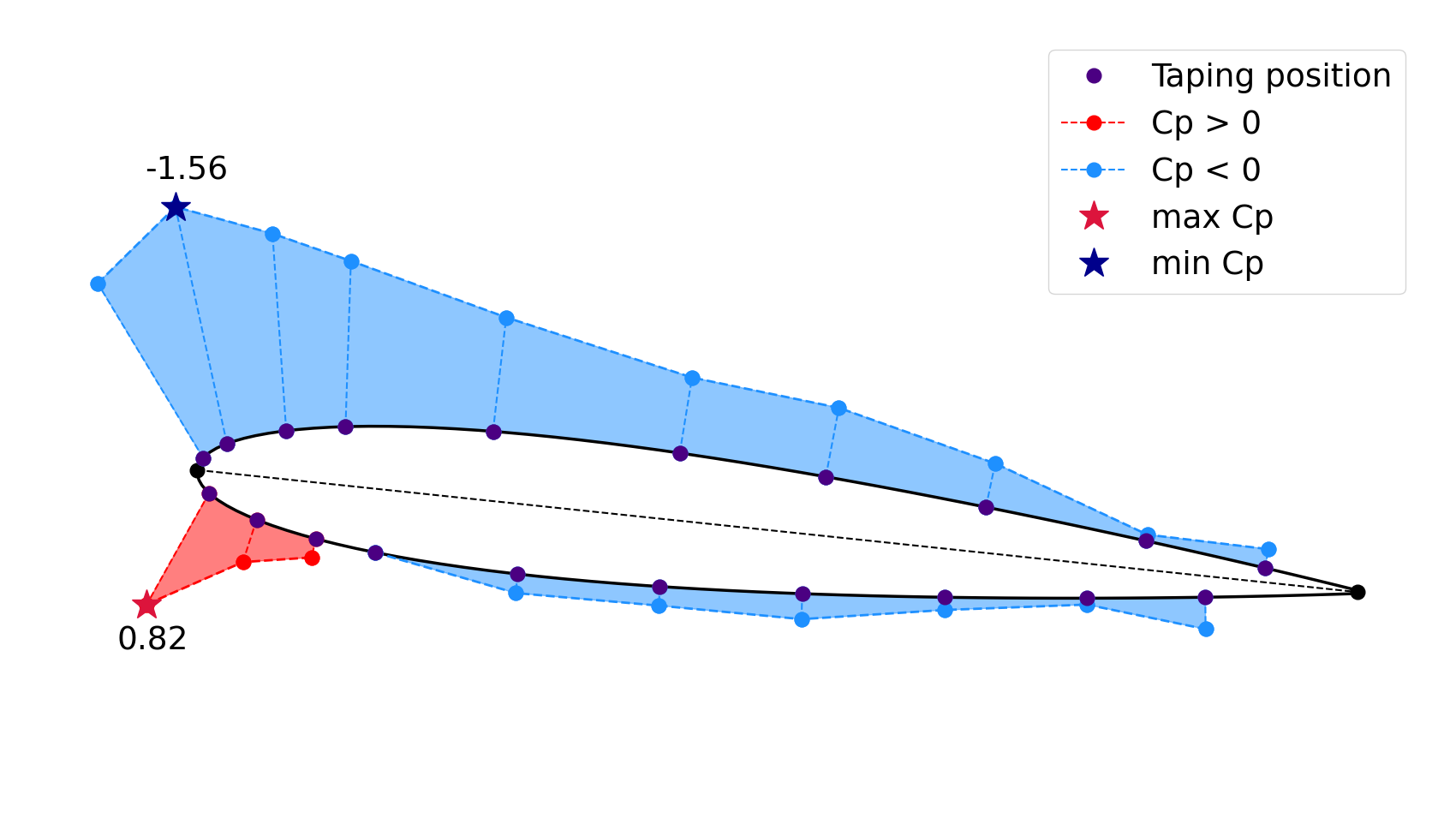}\label{Cp_dis6}}
  \medskip
  \subfloat[AoA = 8°.]{\includegraphics[width=0.48\linewidth]{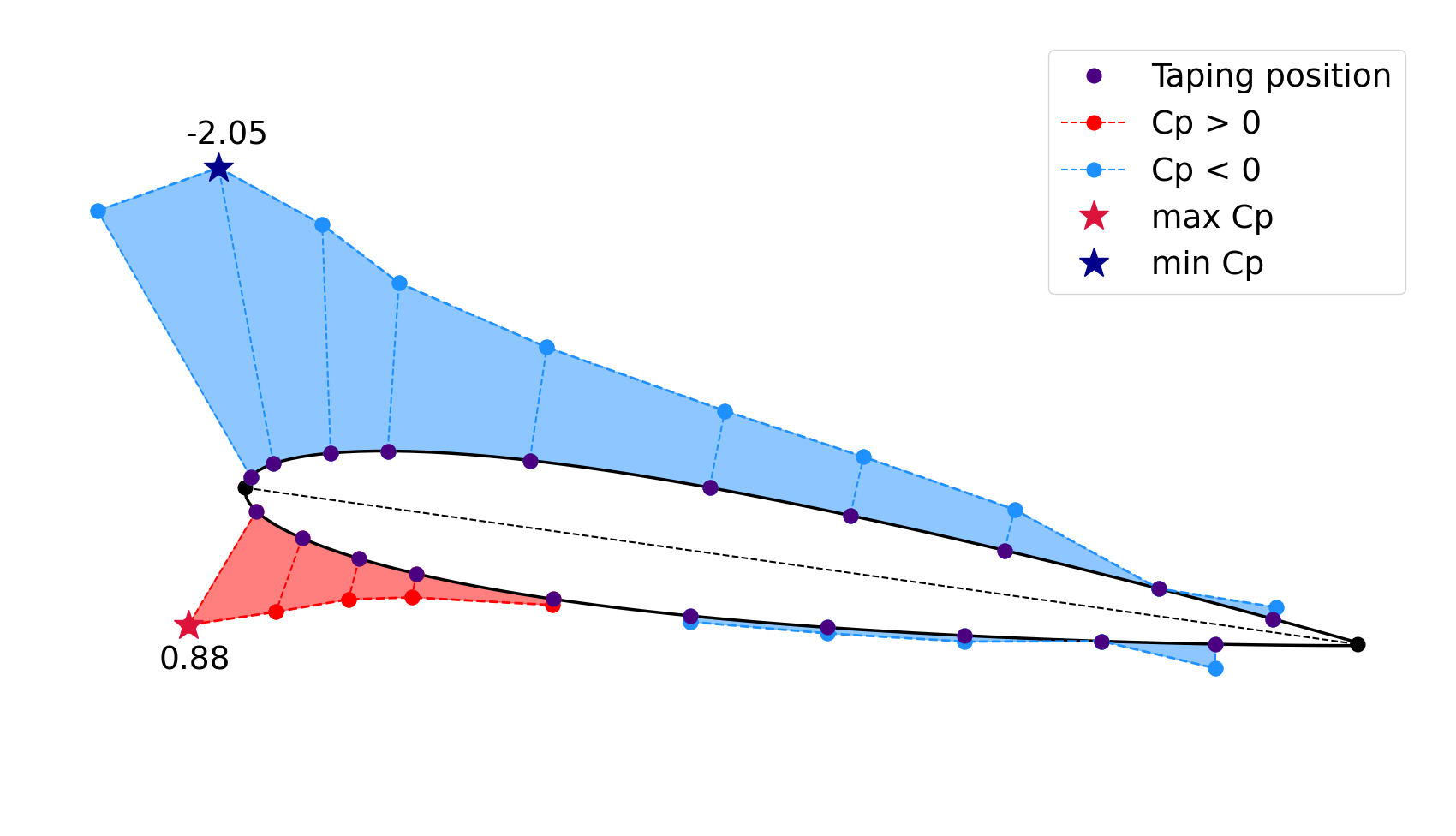}\label{Cp_dis8}}
  \caption{Pressure distribution over NACA0012 airfoil for varying angles of attack where Reynolds number equals Re = 191000. Recorded minimal and maximal values of $C_p$ are given at respective taping positions.} \label{Cp_distribution}
\end{figure}

With the increase in AoA, maximal and minimal values of $C_p$ are rising. The stagnation point is moving downstream, which is in accordance with \citep{Douvi2010}. An additional overview of the $C_p$ distribution for different AoA is given in \cref{Cp_distribution}. In general, with the increase in AoA, pressure coefficient is increasing on the lower surface and dropping on the upper surface.

Disparity in experimental results can be attributed to measurement errors. Due to minor imperfections of the ventilator, diffuser, contractor, honeycomb, and the entire test section, it is easy to assume that fluid flow at the inlet isn't perfectly uniform. Moreover, every measurement tool, as mentioned in \cref{exp}, has a small measurement imperfection, even with a proper calibration. Therefore, experimental results, shown in \cref{Cp_8_graph}, \cref{Cp_Exp} and \cref{Cp_distribution}, will have uncertainties.

Those uncertainties can be quantified through the standard deviation of the measured pressure coefficient. As it was described in \cref{exp}, the fluid flow in the experiment was recorded for 300 seconds with a 0.5 second interval, making a total of 600 measured values. The measured parameters were used to calculate $C_p$ as in \cref{Cp}, but for every time step. Therefore, from a newly generated $C_p$ list, the standard deviation for the pressure coefficient can be calculated using \cref{std}:

\begin{equation}
    \sigma = \sqrt { \frac{ \sum_{i=1}^{n} {(C_{p_i} - \overline{C_p})}^2 }{n-1}}
    \label{std}
\end{equation}

where $\sigma$ represents sample standard deviation, while n is the number of values in the sample. Likewise, $C_{p_i}$ determines $C_p$ of one measurement in a loop with range of 600, while $\overline{C_p}$ is the mean pressure coefficient.

\begin{figure}[H]
  \centering
  \subfloat[AoA = 2°.]{\includegraphics[width=0.48\linewidth]{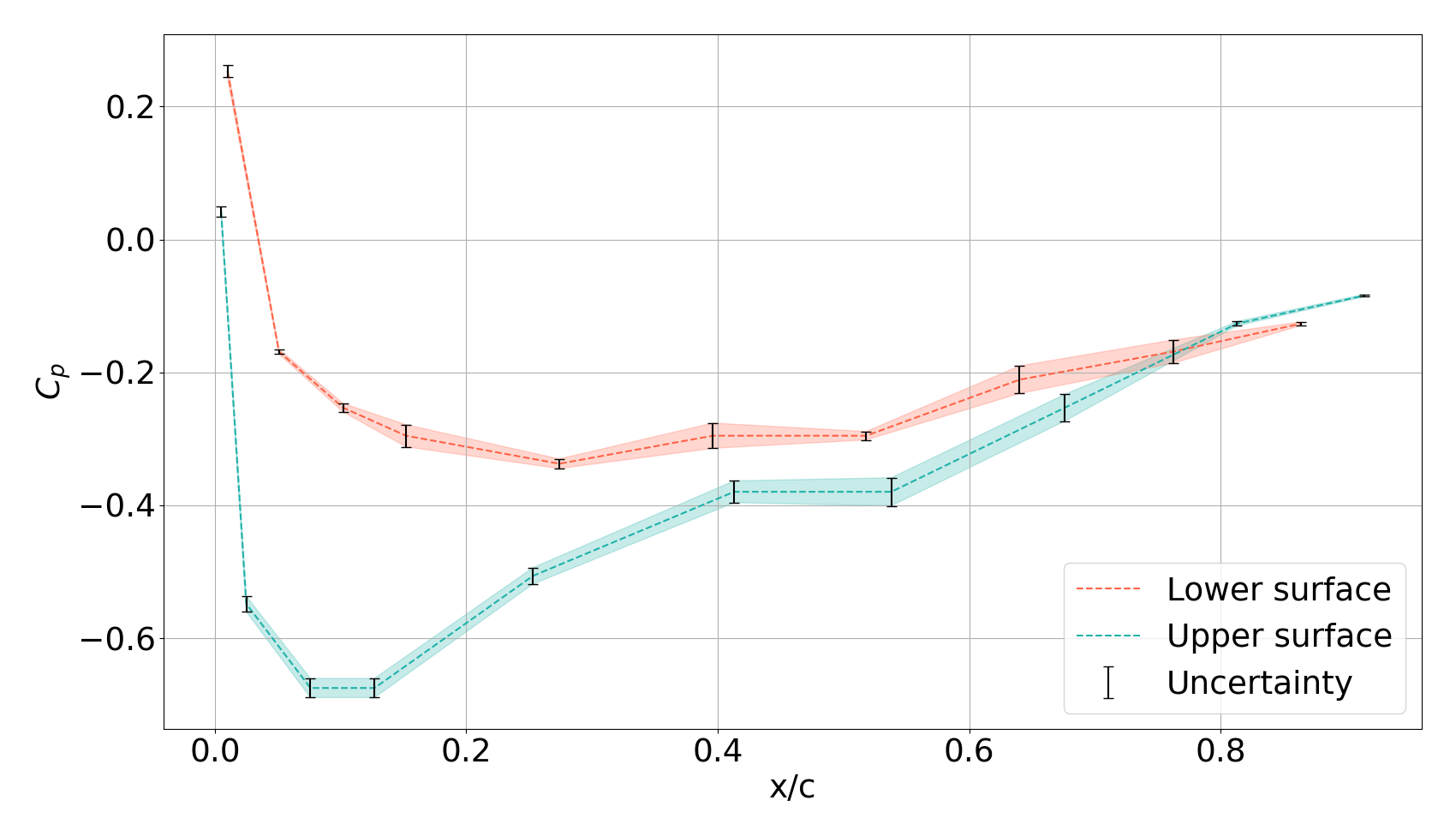}\label{err2}}
  \medskip
  \subfloat[AoA = 4°.]{\includegraphics[width=0.48\linewidth]{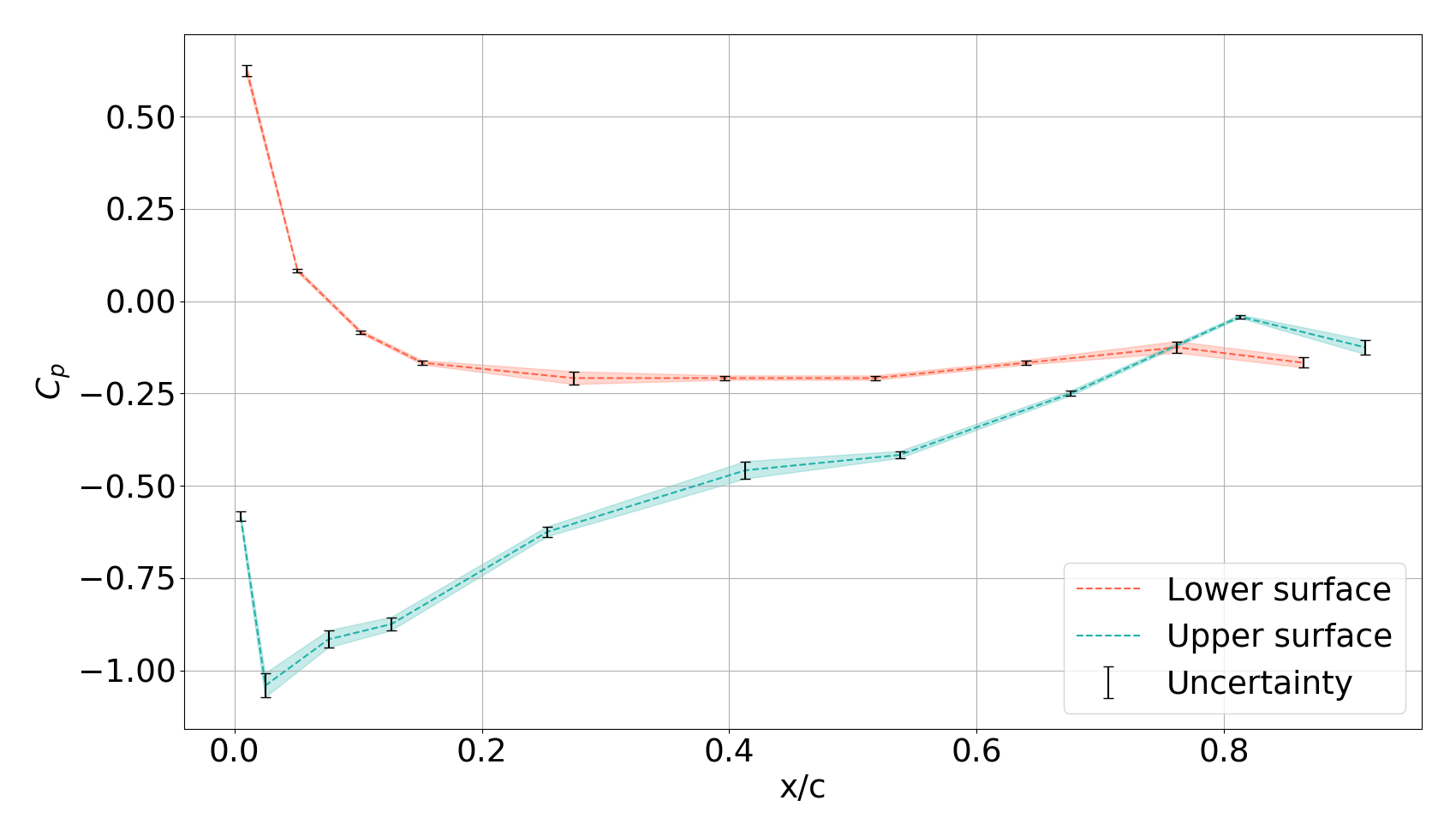}\label{err4}}
  \hfill
  \subfloat[AoA = 6°.]{\includegraphics[width=0.48\linewidth]{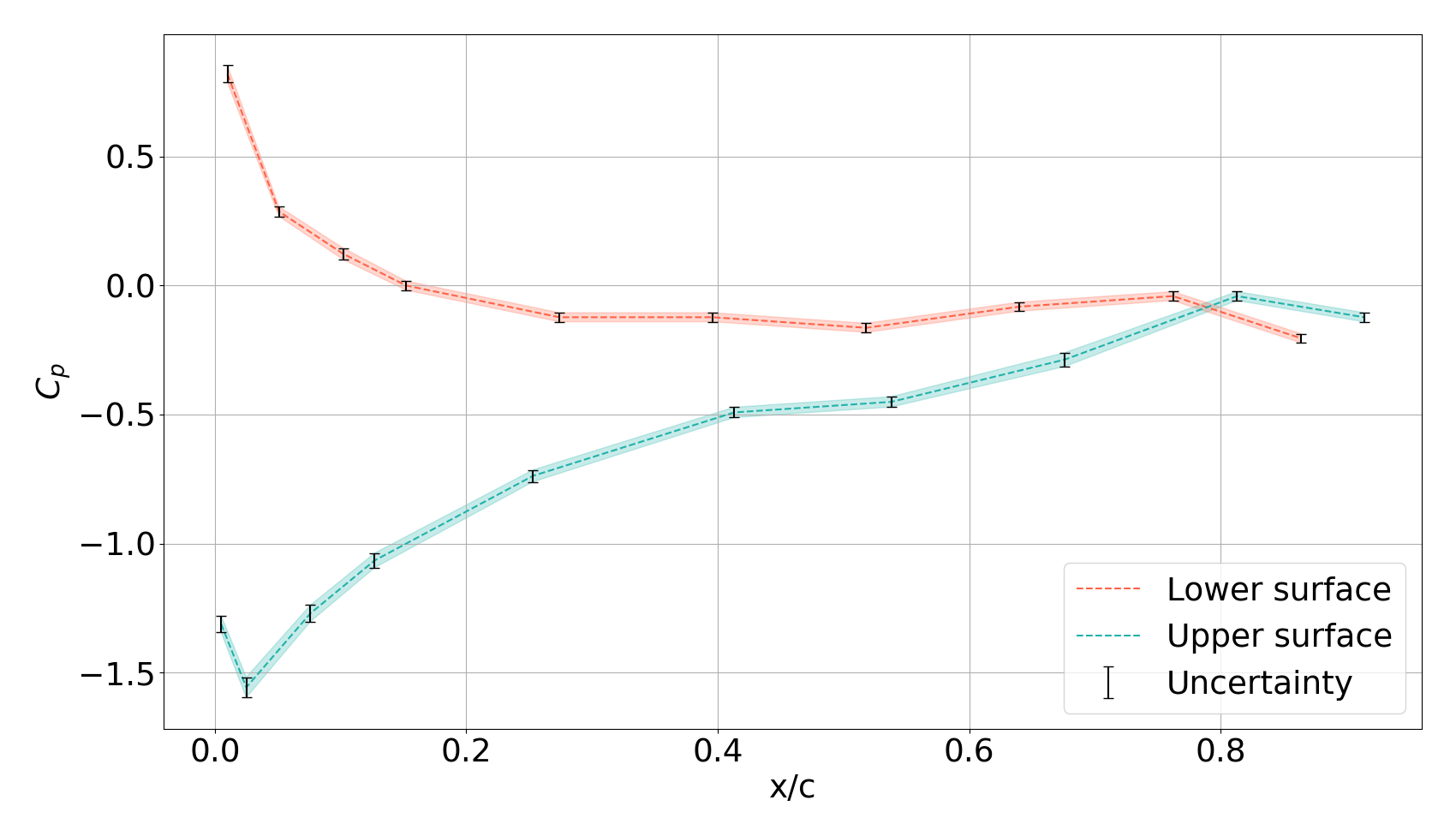}\label{err6}}
  \medskip
  \subfloat[AoA = 8°.]{\includegraphics[width=0.48\linewidth]{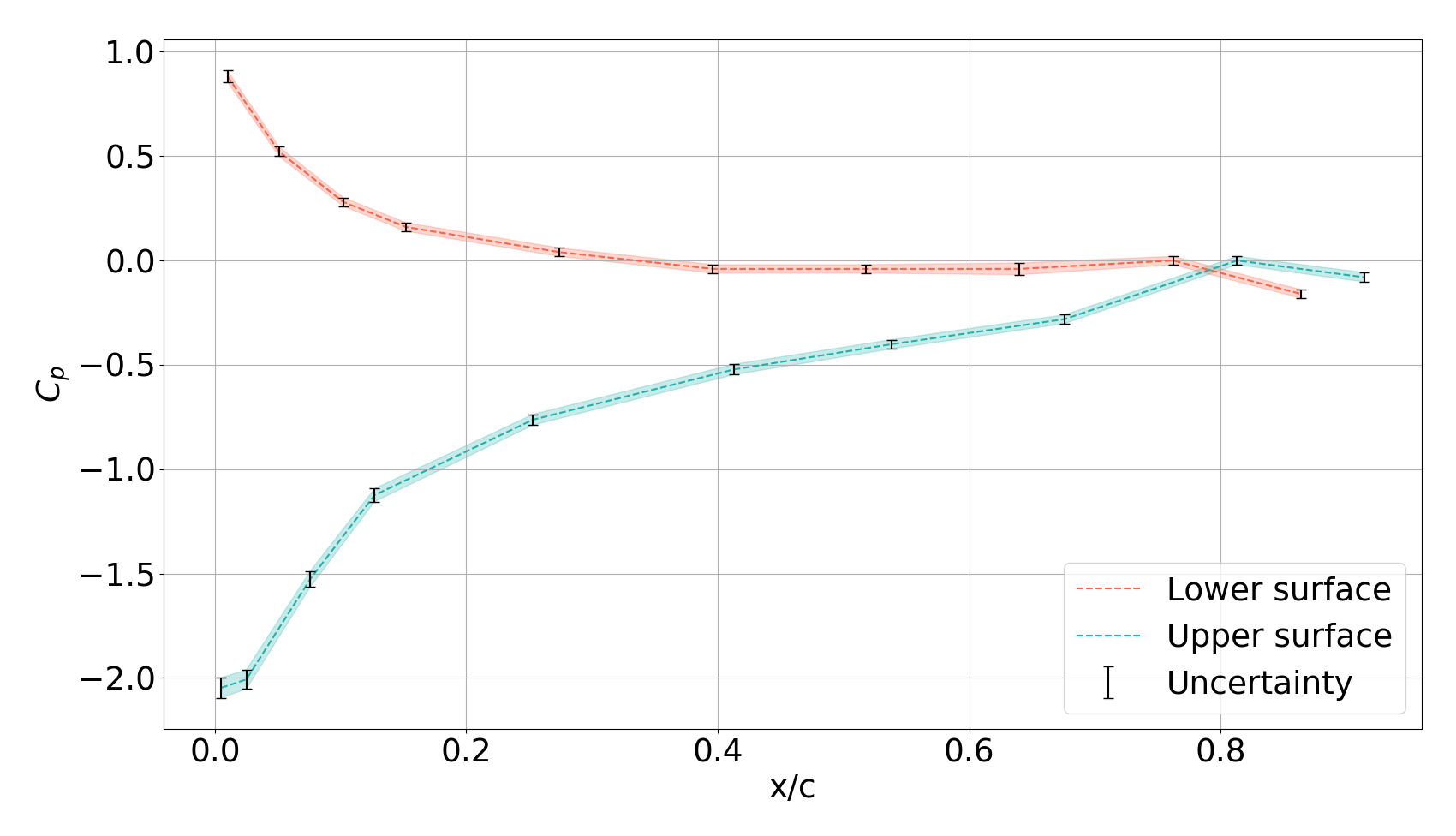}\label{err8}}
  \caption{Experimental results for different AoA with $C_p$ uncertainties.} \label{err}
\end{figure}

Measurement uncertainties calculated for all experiments, i.e. four different AoA cases, are illustrated in \cref{err}.

We can appraise that uncertainties displayed in \cref{err} are pretty small, which suggests that the measurements are acceptable. Similarly with \cref{err}, detailed numerical values are shown in \cref{unc}, where maximal standard deviations are represented for all taping positions on the airfoil, regarding the different AoA.

\begin{table}[H]
\renewcommand{\arraystretch}{1.2}
\caption{Maximum pressure coefficient standard deviation for lower and upper surface with their taping position.\label{unc}}
\newcolumntype{C}{>{\centering\arraybackslash}X}
\begin{tabularx}{\textwidth}{CCCCC}
\toprule
\textbf{Angle of attack}	& \textbf{\bm{$STD_{max}$} upper surface}	& \textbf{Taping position} & \textbf{\bm{$STD_{min}$} lower surface} & \textbf{Taping position}\\
\midrule
2°		& 0.0211		& 13       & 0.02068     & 16\\
4°		& 0.03276		& 3        & 0.01733     & 10\\
6°		& 0.03951		& 3        & 0.03309     & 2\\
8°		& 0.04793		& 1        & 0.02803     & 16\\
\bottomrule
\end{tabularx}
\end{table}

Based on \cref{unc}, we can conclude that the biggest standard deviation for the upper surface appears for the 8° AoA at tapping position 1. On the other hand, for the lower surface, the biggest deviation is at tapping position 2 for the 6° AoA. Therefore, all measurement uncertainties have a standard deviation error below 5\%. We can summarize that further comparisons with numerical results can be carried out with averaged experimental results.


\subsection{Mesh assessment for LBM simulations}
    \label{mesh_ass}
Mesh independence study is essential for ensuring result congruence under different conditions. Result consistency regardless of the numerical mesh minimizes potential grid-related numerical errors. In this study, pressure coefficient and velocity values have been evaluated and compared for cases with varying mesh sizes. Initially, four different numerical grids, introduced in \cref{Num_setup}, have been generated and analysed with regards to $C_p$ distribution. Presented results are LBM regression curves for lower and upper surfaces. Only the test case where AoA equals 8° has been considered for mesh independence analysis. As eddies are more frequent and the overall turbulence of the flow is more pronounced, it is appropriate to assume that if valid for noted AoA, mesh and methodology would be valid for the remaining AoA cases as well.

\begin{figure}[H]
  \centering
  \subfloat[Lower surface.]{\includegraphics[width=0.48\linewidth]{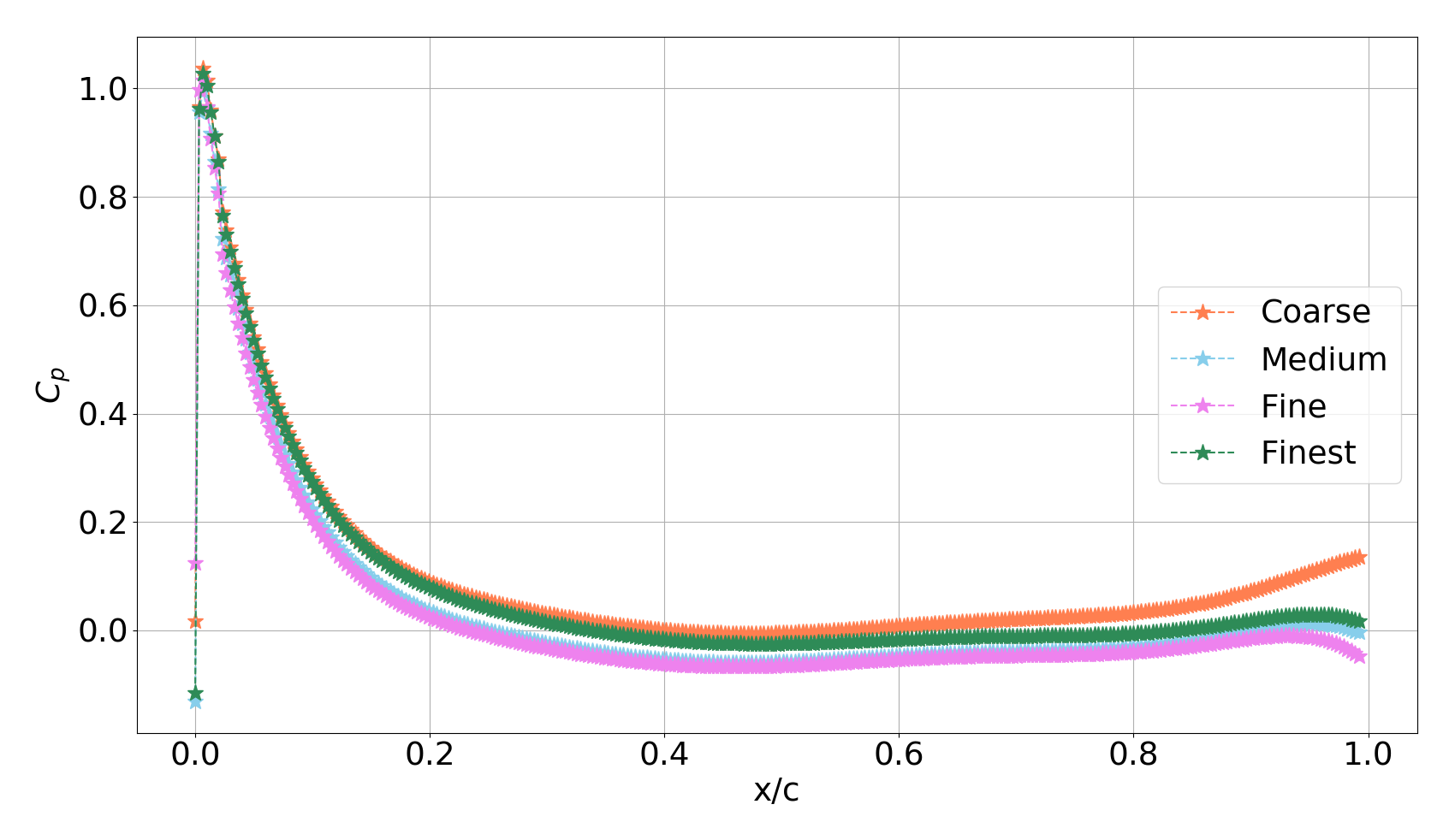}\label{Low_mesh}}
  \hfill
  \subfloat[Upper surface.]{\includegraphics[width=0.48\linewidth]{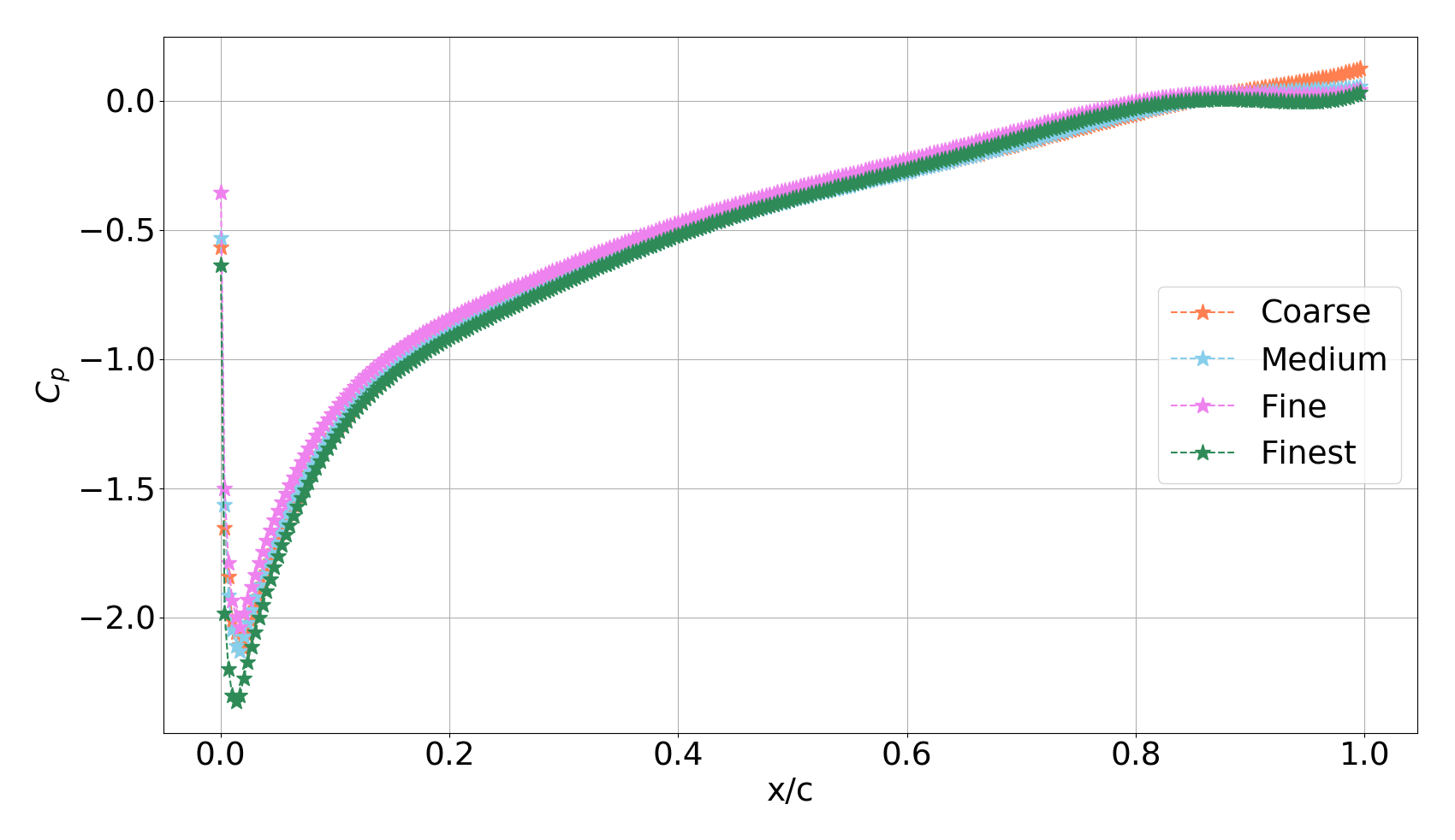}\label{Up_mesh}}
  \caption{Results of the mesh independence for 8° AoA of NACA0012 airfoil. Four different numerical grids are considered. (a) results on lower surface, (b) results on upper surface.} \label{mesh}
\end{figure}

\cref{mesh} suggests a relative equivalence between the results for different grids. On the lower surface (\cref{Low_mesh}), results near the trailing edge for coarsest mesh are diverging. Remaining grids exhibit consistent and similar trends. Analogously, results for the upper surface presented in \cref{Up_mesh} suggest general similarity between trends. Clearly, near the trailing edge, coarsest mesh deviates the most. At the leading edge, the finest mesh achieves the lowest $C_p$ values. This anomaly can be attributed to the extensive grid refinements near the wall which enable detailed capture of the flows specifics.

Absolute maximal values for meshes represented in \cref{mesh} are additionally quantified and given in \cref{mesh_ind_T1} as maximal $C_p$ for lower surface and minimal $C_p$ for upper surface. As noted in \cref{exp_res}, maximal $C_p$ should be one. The coarsest grid provides the worst result, with a 3.68\% disparity in maximal $C_p$ value. Remaining results are within 3\% from the expected value of one. When assessing minimal $C_p$, results on coarse, medium and fine grid are comparable. $C_p$ for the finest mesh differs, which can be attributed to numerical errors on an exceedingly refined grid.

\begin{table}[H]
\renewcommand{\arraystretch}{1.2}
\caption{Maximal and minimal $C_p$ values with number of total fluid voxels for different mesh type cases.\label{mesh_ind_T1}}
\newcolumntype{C}{>{\centering\arraybackslash}X}
\centering
\begin{tabularx}{.9\linewidth}{CCCC}
\toprule
\textbf{\makecell{Mesh type \\ case}} & \textbf{\makecell{Total fluid \\ voxels [$\times 10^6$]}} & \textbf{\makecell{Max $C_p$}} & \textbf{\makecell{Min $C_p$}}\\
  \midrule
    Coarse      & 3.66		& 1.03681          & -2.12141\\
    Medium      & 8.01      & 0.99453          & -2.12935\\
    Fine        & 12.69     & 1.01443          & -2.03686\\
    Finest      & 17.02     & 1.02795          & -2.32405\\
  \bottomrule
\end{tabularx}
\end{table}

In order to properly compare results from \cref{mesh}, RMSE has been calculated according to \cref{RMSE},

\begin{equation}
    RMSE = \sqrt{\sum_{i=1}^{n}\frac{[(C_{p})_{i} - (C_{p})_{i+1}]^2}{n}}
    \label{RMSE}
\end{equation}

where $i$ represents different mesh cases. Accordingly, the discrepancy between results for different mesh types is given in \cref{mesh_ind_T2}. Lower surface RMSE has the worst result for the coarsest mesh, because of the inconsistent trailing edge that can be seen in \cref{Low_mesh}. Regarding to upper surface, finest mesh differs from others. Reason is a finer grid that surrounds the airfoil much closely and more precisely, which gives a more detailed view of the $C_p$ distribution, especially around the leading edge, that is also portrayed in \cref{Up_mesh}. Overall, the results are mostly in congruence and do not vary significantly, hence, with regards to the pressure coefficient value, grid consistency has been demonstrated.

\begin{table}[H]
\renewcommand{\arraystretch}{1.2}
\caption{$C_p$ RMSE comparison between different mesh type cases for both lower and upper surface.\label{mesh_ind_T2}}
\newcolumntype{C}{>{\centering\arraybackslash}X}
\centering
\begin{tabularx}{0.8\linewidth}{CCCC}
\toprule
  \textbf{\makecell{Mesh type \\ case}} & \textbf{\makecell{RMSE \\ Lower surface}} & \textbf{\makecell{RMSE \\ Upper surface}}\\
  \midrule
    Coarse-Medium		  & 0.06272          & 0.0187\\
    Medium-Fine           & 0.02018          & 0.04936\\
    Fine-Finest           & 0.04883          & 0.08432\\
  \bottomrule
\end{tabularx}
\end{table}

\begin{figure}[H]
      \centering
      \includegraphics[width=0.99\columnwidth]{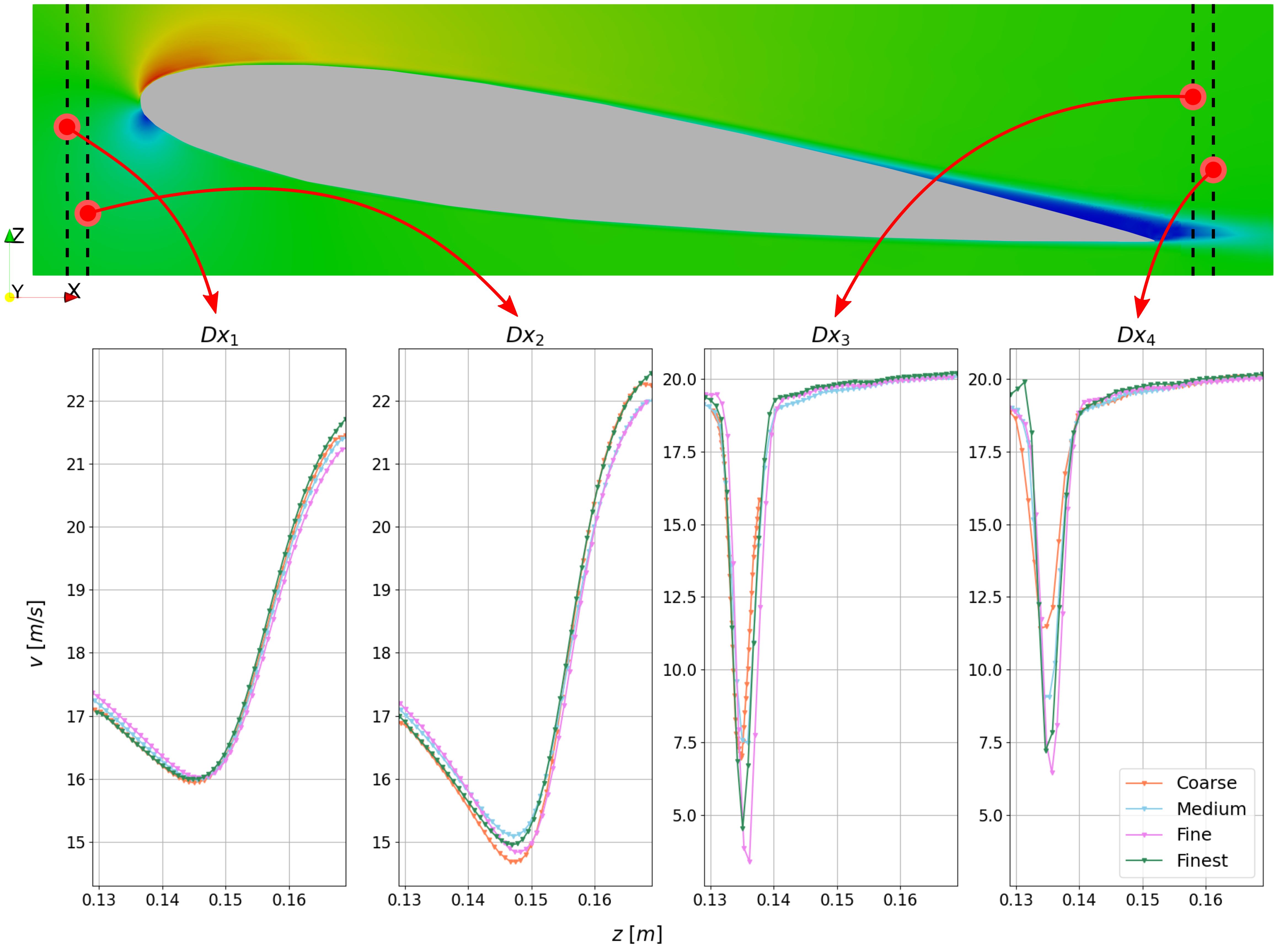}
      \caption{Velocity profiles obtained at different cross sections for the 8° AoA case on different grids. Noted YZ cross sections ($Dx_1$ to $Dx_4$) are located at X=0.182m, X=0.185m, X=0.348m, and X=0.351m, respectively. Graphs display velocity magnitude at each cross section.\label{mesh_ind_vel}}
\end{figure}

With regards to the velocity distribution around the airfoil, assessment has been conducted by evaluating velocity magnitude at the centerline of several YZ cross sections. Obtained profiles and considered locations are shown in \cref{mesh_ind_vel}. Top part of \cref{mesh_ind_vel} represents a segment of the computational domain in XZ plane. Consequently, the contour start from 0.177m and finished at 0.36m, in X direction, while in Z direction it starts at 0.129m and ends at 0.169m. Dashed lines represent cross sections, while velocity distribution through the centerline of the YZ plane is illustrated in the lower part. Graphs for sections $Dx_1$ and $Dx_2$ indicate that upstream of the airfoil, due to the influence of the uniform inflow, values of velocity for different grids are mostly consistent. The second graph ($Dx_2$) additionally reveals small deviations in the coarsest mesh near the leading edge. Values in the wake region tend to fluctuate as the velocity drops. These differences are a direct consequence of insufficient mesh density (refinement) in this region which leads to inability to properly capture complex flow structures. Nevertheless, a similar shape for different meshes indicates good mesh independence quality, even in the turbulence region. Besides, downstream of the airfoil, velocity approaches uniformity observed at upstream locations.

Based on the conducted assessment, several conclusions can be drawn. Coarse mesh is inadequate near the leading and trailing edge and should be avoided. Remaining grids are mostly appropriate and provide similar results for both pressure coefficient and velocity. In order to account for any potential mesh-associated errors, the finest grid is hence chosen for future analysis, as it introduces grid refinements which should be adequate to properly characterize the airfoil and allow detailed LES simulation.


\subsection{Numerical results}
    \label{num_res}
\begin{figure}[H]
      \centering
      \includegraphics[width=0.55\columnwidth]{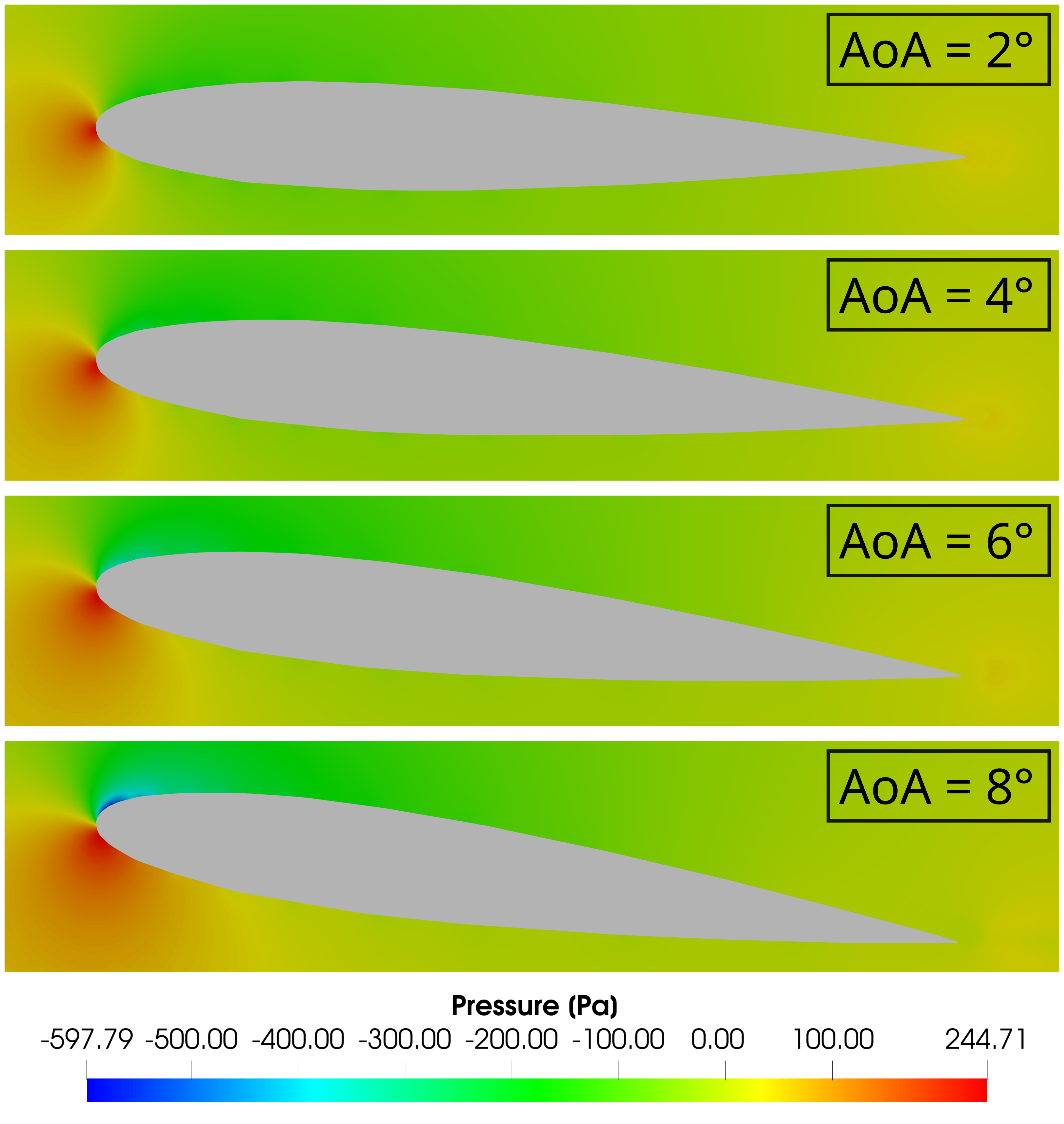}
      \caption{Pressure contours at cross section y=0 for different AoA.\label{p}}
\end{figure}

Numerical setup described in \cref{Num_setup} is henceforth employed on chosen mesh for all AoA test cases. As expected, results presented in \cref{p} suggest that with the increase in AoA, pressure gradient increases as well. The airfoil profile experiences a favorable pressure gradient in front of the leading edge while further downstream, adverse pressure gradient is noted.

\begin{figure}[H]
      \centering
      \includegraphics[width=0.9\columnwidth]{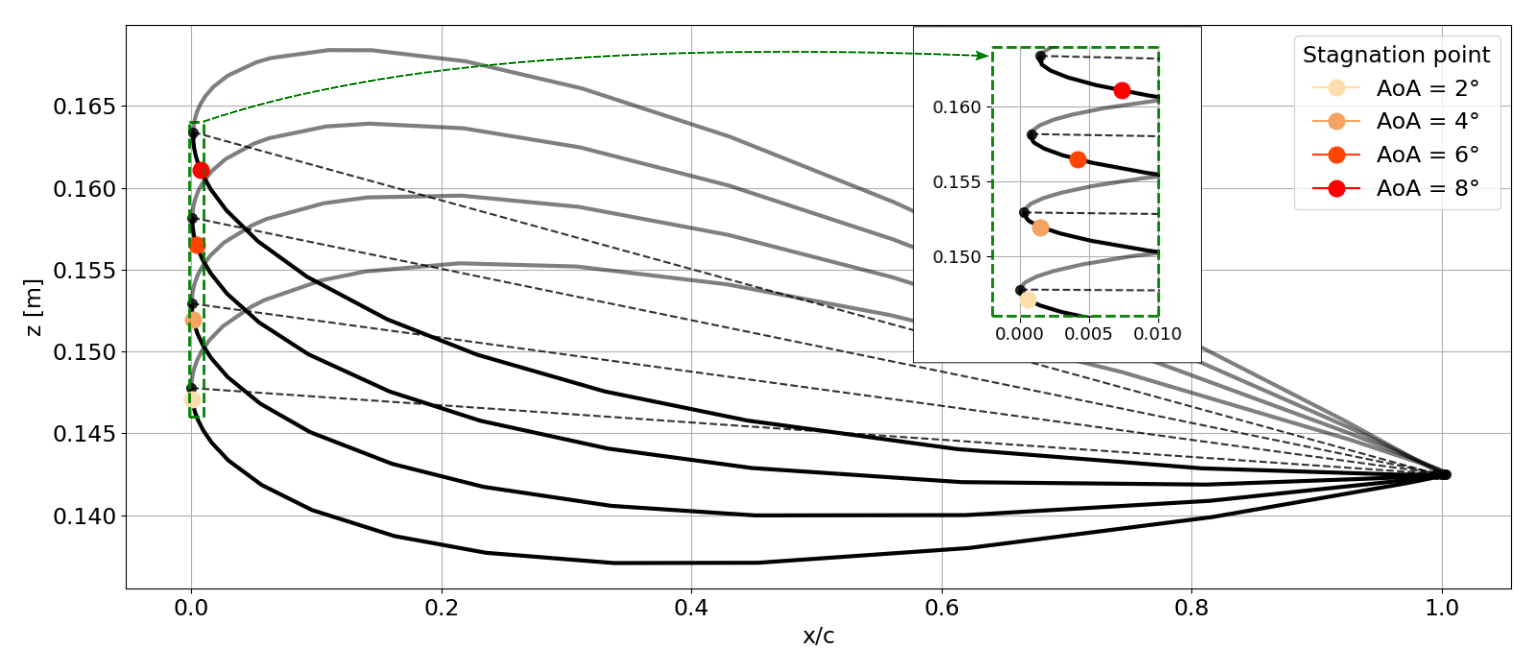}
      \caption{Stagnation point distribution for different AoA cases. Lower surface is shown in black, while the grey color is representing upper surface.\label{stag}}
\end{figure}

Observed pressure difference for 8° AoA is the largest, with the stagnation point moving downstream and towards the lower surface in comparison to the lower AoA cases, which is highlighted in \cref{stag}.

\begin{figure}[H]
      \centering
      \includegraphics[width=0.55\columnwidth]{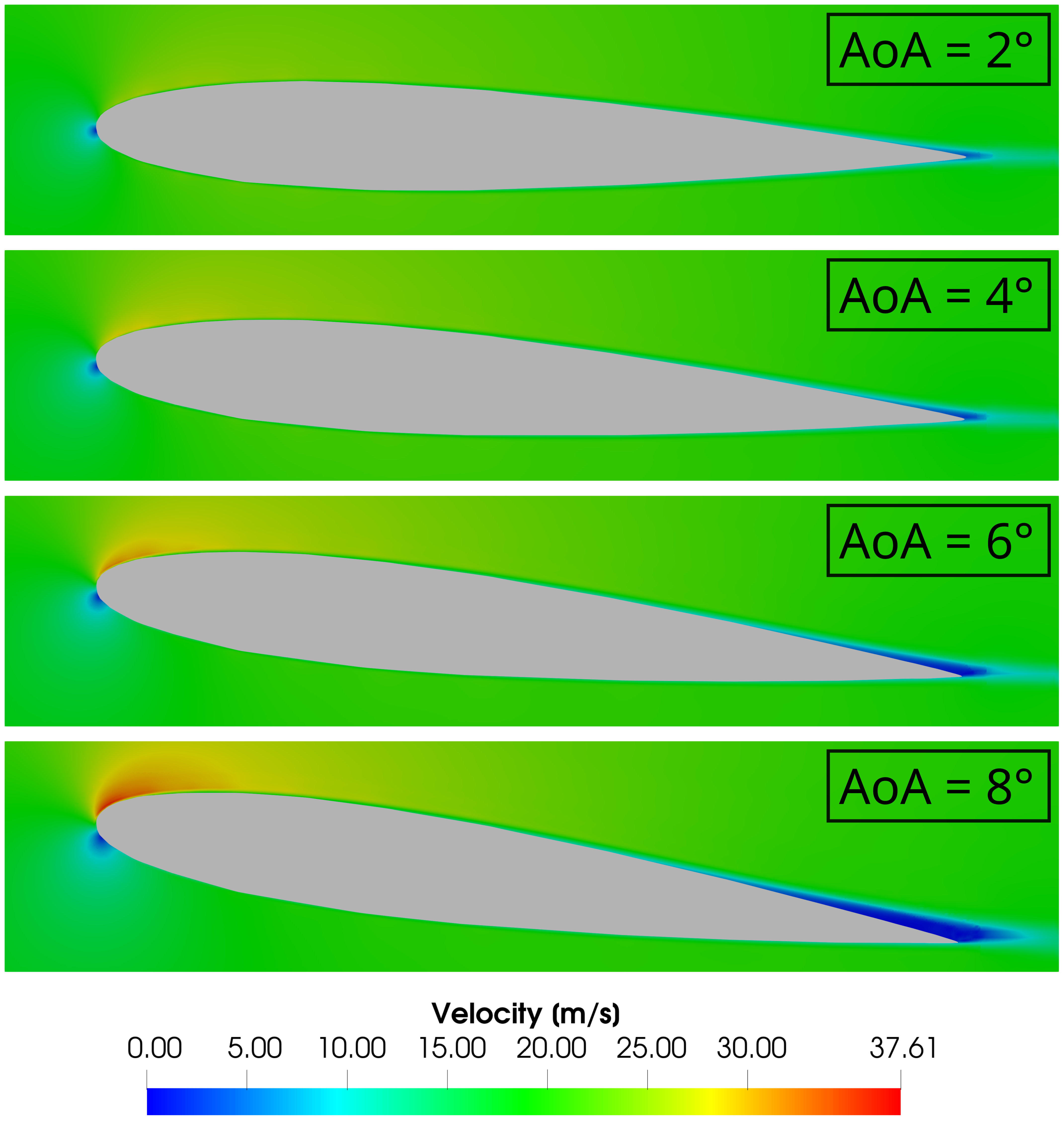}
      \caption{Velocity contours at cross section y=0 for different AoA.\label{v}}
\end{figure}

As \cref{v} shows, the velocity drop is closer to the leading edge for larger AoA values. Consequently, a separation bubble, with reverse flow, can occur in that region. Depending on the slope, the aforementioned bubble can encompass the whole upper surface, which in turn leads to stall. For our four cases, stall didn’t occur, because of adequate AoA and velocity. Stall will be induced for larger AoA and Reynolds numbers as concluded by \citet{Almohammadi2022}, where Re was 360000 and AoA 10°. On the lower surface, the boundary layer remained attached to the airfoil, but it separated on the upper surface. Separation point depends on the velocity and the AoA. Finally, with the increase in AoA, the wake region behind the airfoils is increasingly turbulent. Velocity on the upper surface is evidently dependant on the AoA as it increases with it, however, near the trailing edge, it effectively returns to zero.

\begin{figure}[H]
  \centering
  \subfloat[AoA = 2°.]{\includegraphics[width=0.48\linewidth]{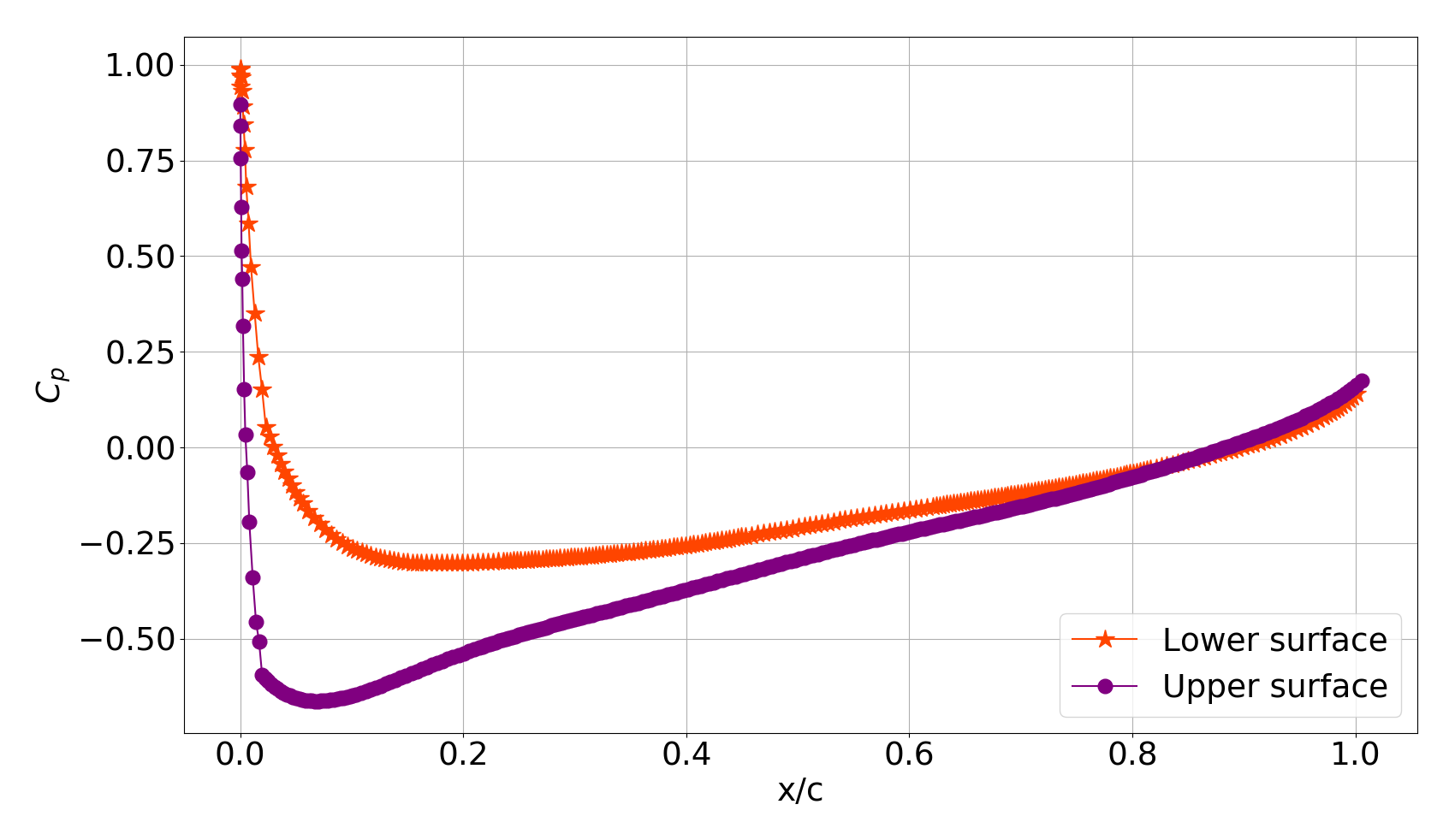}\label{LBM_Cp2}}
  \medskip
  \subfloat[AoA = 4°.]{\includegraphics[width=0.48\linewidth]{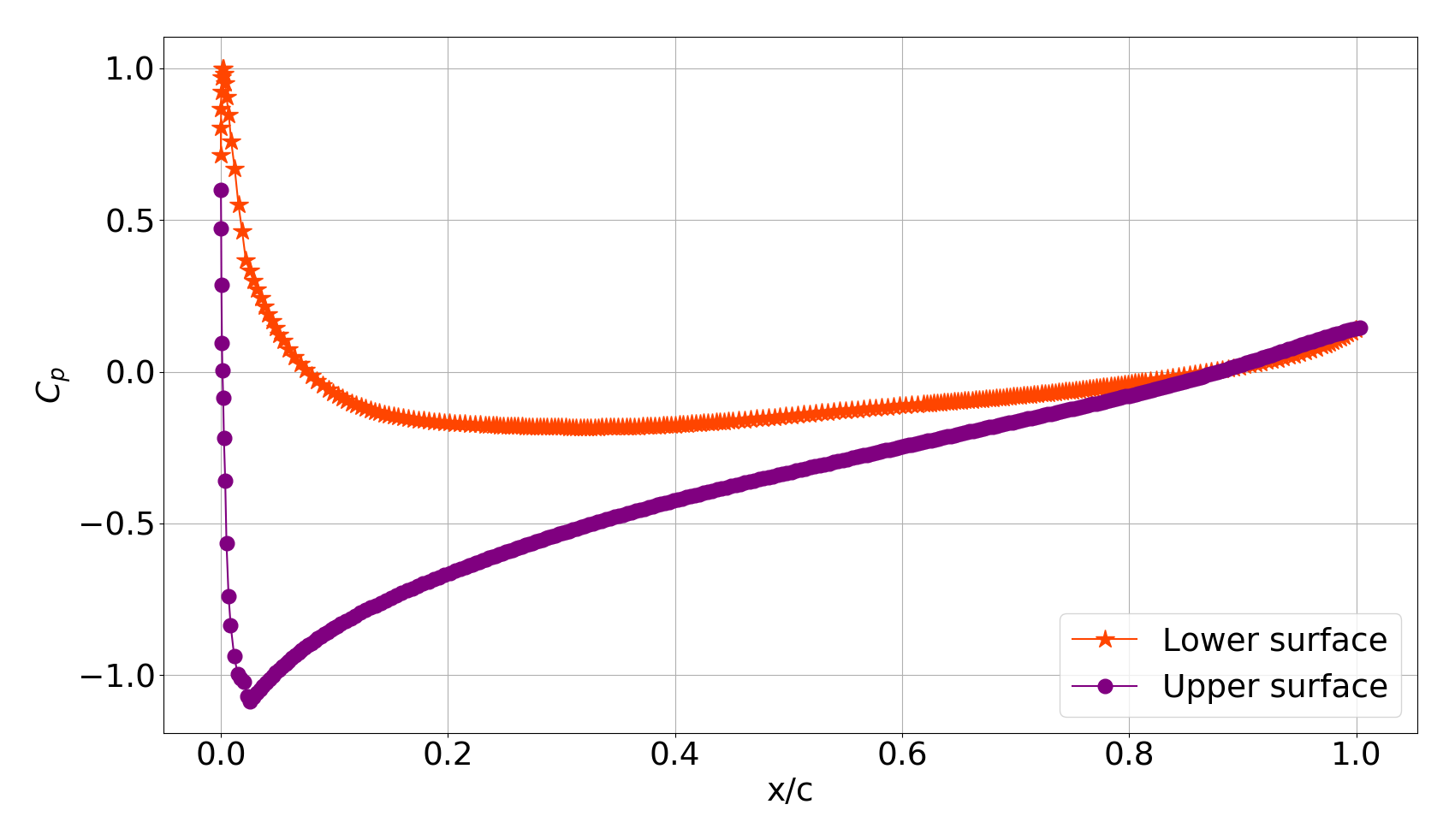}\label{LBM_Cp4}}
  \hfill
  \subfloat[AoA = 6°.]{\includegraphics[width=0.48\linewidth]{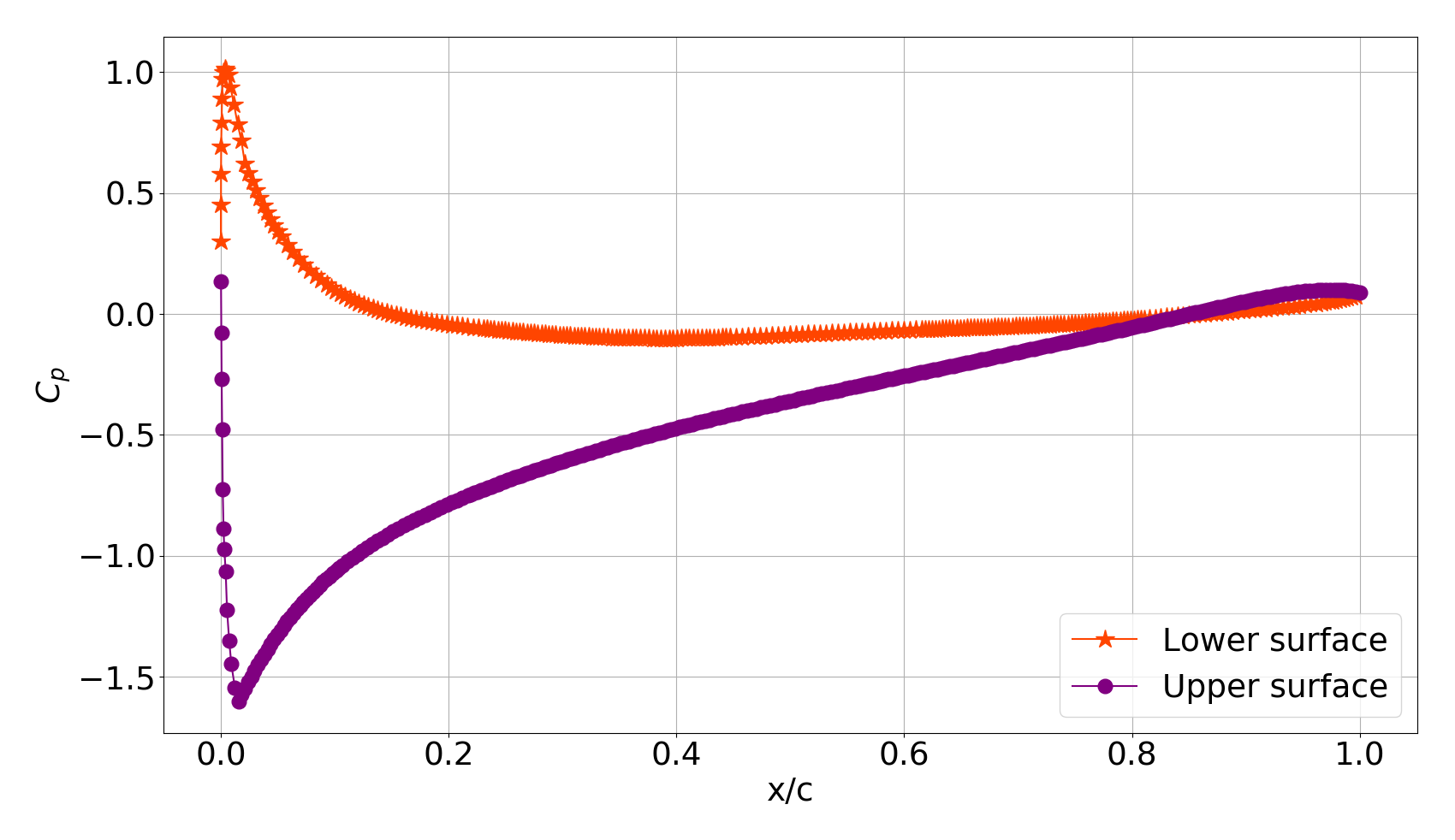}\label{LBM_Cp6}}
  \medskip
  \subfloat[AoA = 8°.]{\includegraphics[width=0.48\linewidth]{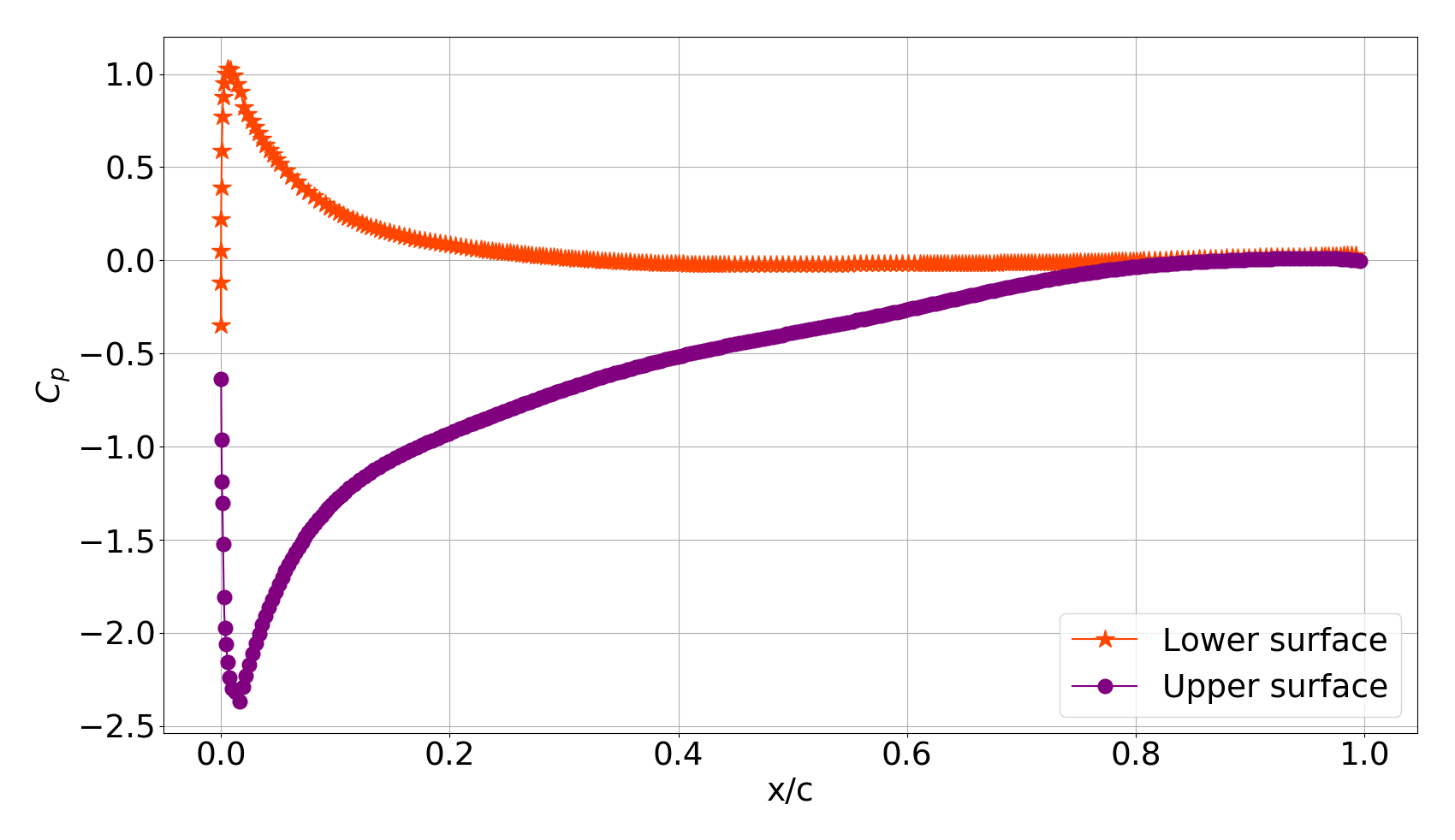}\label{LBM_Cp8}}
  \caption{$C_p$ distribution around NACA0012 airfoil for the finest mesh and different AoA obtained using LBM.}
  \label{LBM_Cp}
\end{figure}

Overall, we can conclude that LBM-LES approach catches the flow around airfoil pretty accurately. \cref{LBM_Cp} displays $C_p$ distribution along the chord. All maximum values are near $C_{p_{max}}$ = 1. The largest deviation of 2.8\% is for 8° AoA, while all other cases have deviation error, for maximal $C_p$, below 1.5\%. Absolute values of the $C_p$ rise with the AoA and are largest for 8 AoA.

With regards to stability check, convergence is ensured at every time step and additionally validated by assessing consistency of drag and lift coefficients at every time step. Despite apparent numerical suitability, the obtained results will henceforth be compared to the experimental results presented in \cref{exp_res}.

\begin{figure}[H]
  \centering
  \subfloat[AoA = 2°.]{\includegraphics[width=0.48\linewidth]{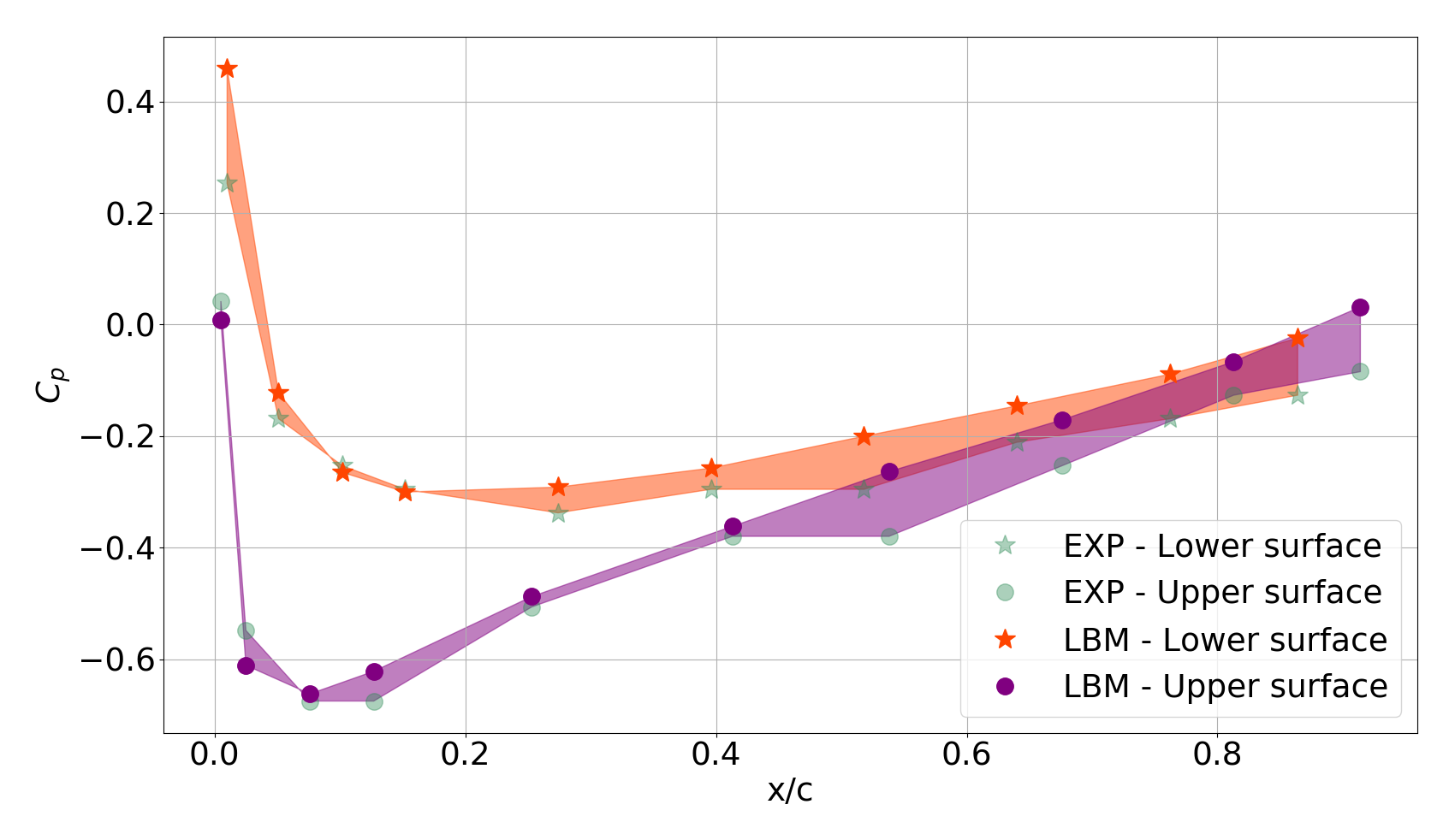}\label{COM_Cp2}}
  \medskip
  \subfloat[AoA = 4°.]{\includegraphics[width=0.48\linewidth]{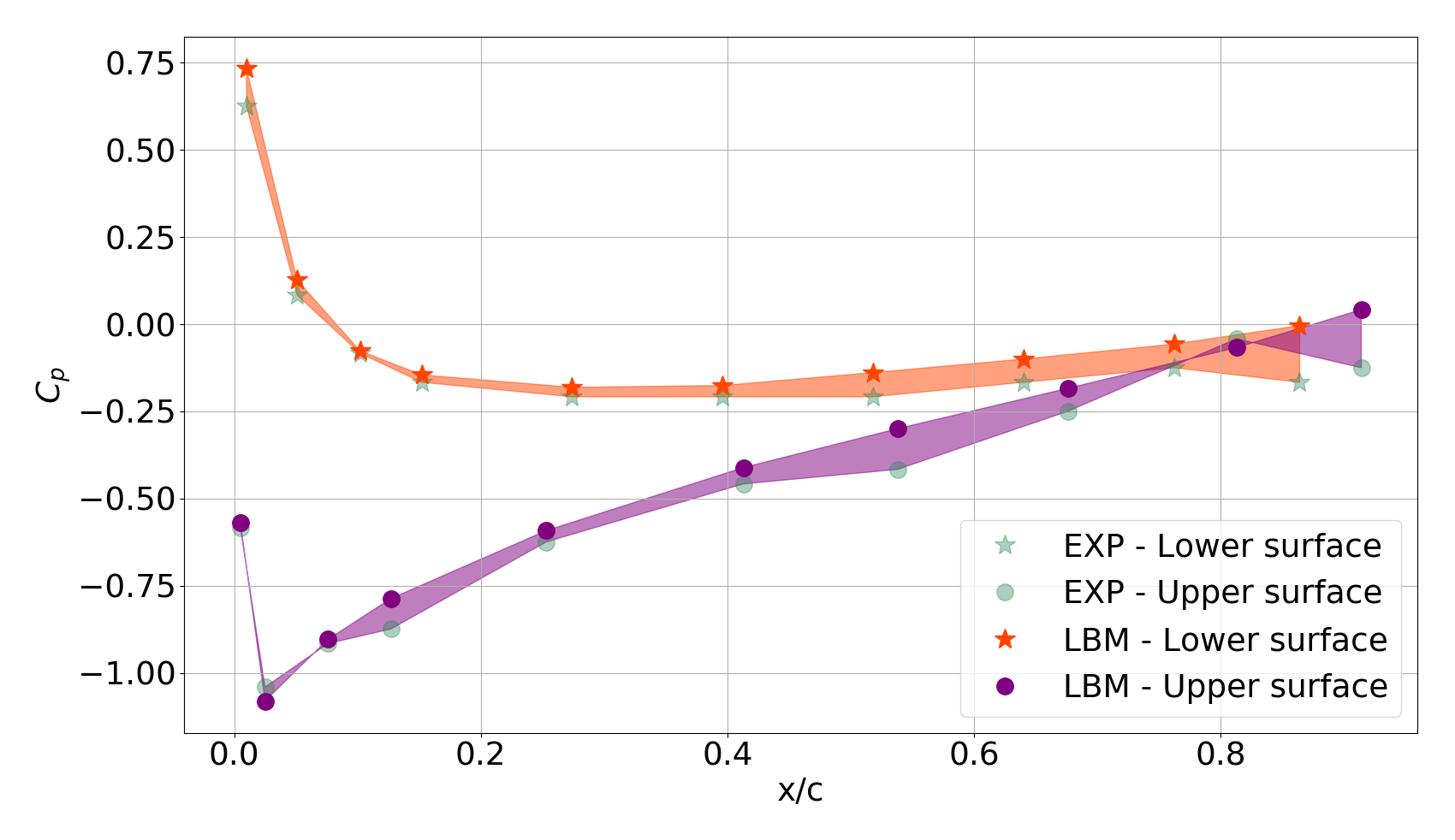}\label{COM_Cp4}}
  \hfill
  \subfloat[AoA = 6°.]{\includegraphics[width=0.48\linewidth]{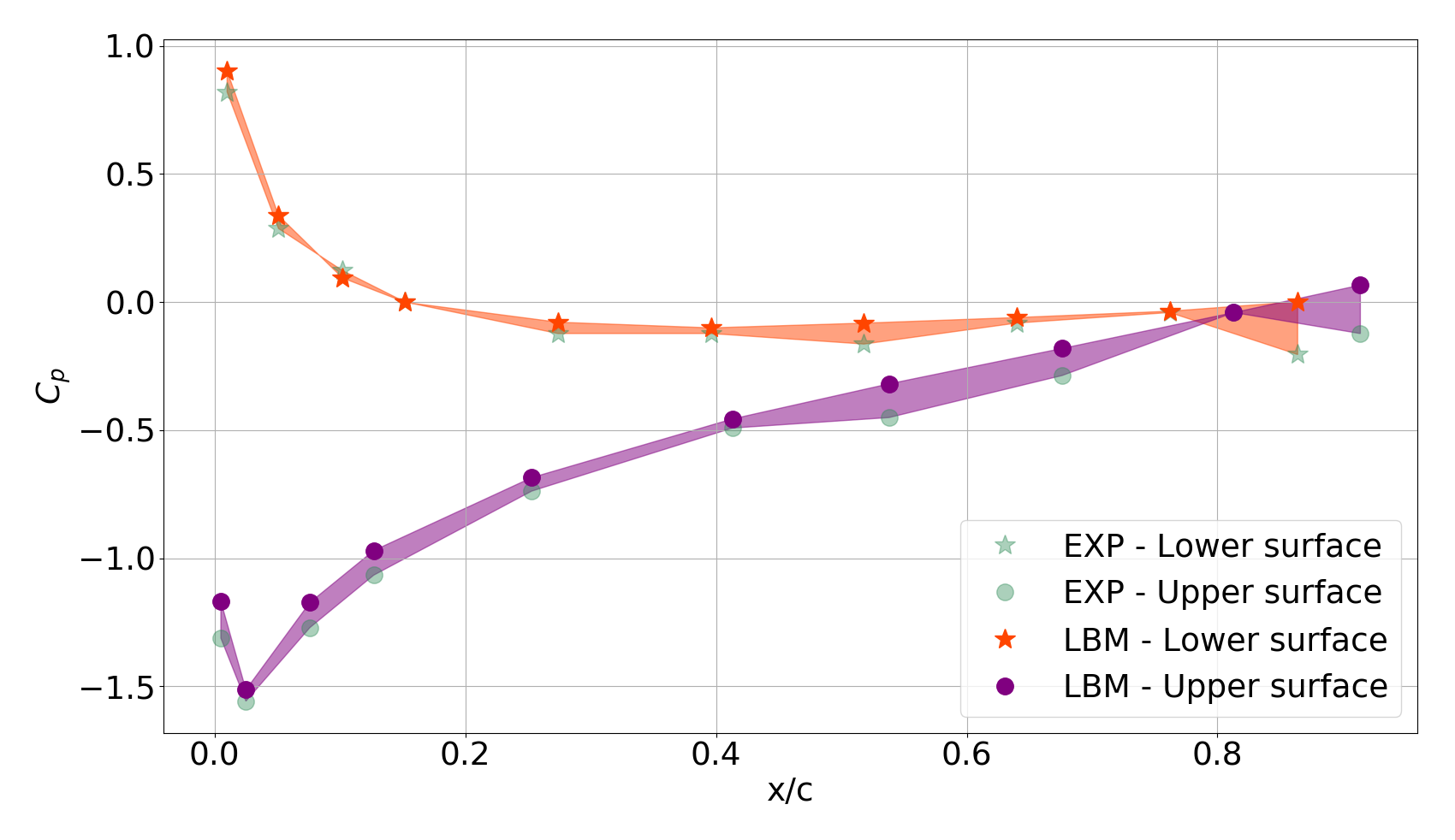}\label{COM_Cp6}}
  \medskip
  \subfloat[AoA = 8°.]{\includegraphics[width=0.48\linewidth]{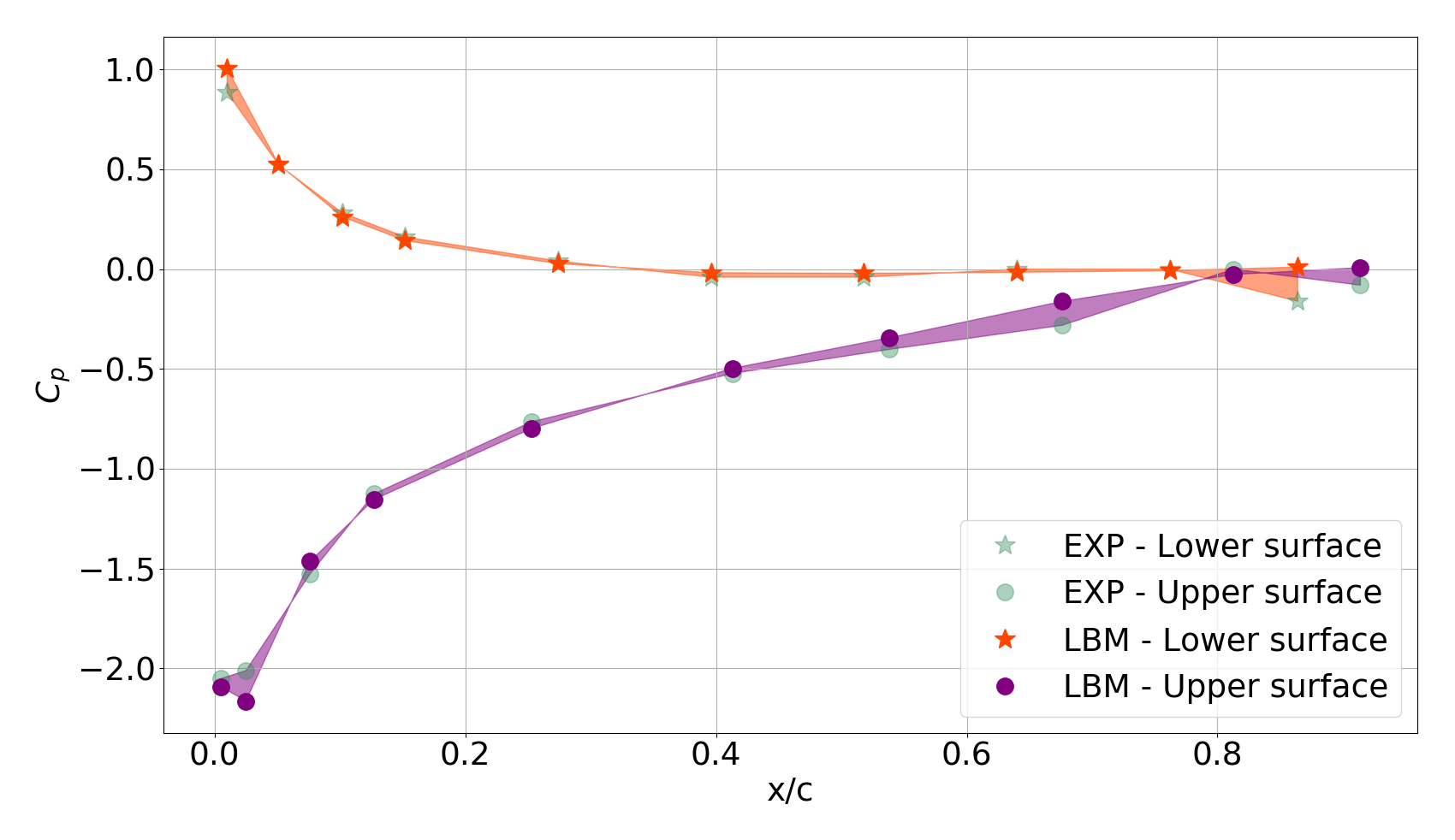}\label{COM_Cp8}}
  \caption{Comparison between LBM results obtained on finest grid and experimental data. Graphs show $C_p$ values for different AoA of the NACA0012 airfoil.}
  \label{COM_Cp}
\end{figure}

\cref{COM_Cp} shows the differences between numerical results and experimental data. Filled regions outline deviations between these sets. Comparison is conducted for 20 experimental measurement points. Good correlations in results can be seen, especially for the \cref{COM_Cp8}. Results for lower surface 8° AoA case show good agreement with the experimental data and overall surpass all other cases in term of accuracy. Moreover, if first and last tapping position on the lower surface (Position 2 and 20) were neglected, the RMSE would drop from 0.068 to 0.014. Likewise, the upper surface is in fine correlation with the experiment, although it has a larger deviation at position 3, which is the absolute maximum value around the airfoil. If \cref{err} is observed, it can be concluded that larger experimental uncertainties exist for positions 1 and 3. Tapping position numbering has been introduced in \cref{exp}, \cref{airfoil}.

Considering 6° AoA case, the largest errors are near the trailing edge, at positions 19 and 20, as is shown in \cref{COM_Cp6}. Disparity is noticeable at the leading edge of the upper surface. However, results match at the lower surface, while LBM results on the upper surface are reduced compared to the experimental one. \cref{COM_Cp4} show 4° AoA case. Deviations at the trailing edge are analogous to the 6° AoA case. Tapping position 2 shows smaller divergence on lower surface, while all other positions have similar and smaller error. In addition to upper surface, deviations are consistent through the airfoil. Interestingly, on the leading edge, \cref{COM_Cp4} shows great data correlation, which proves absolute maximal values as relevant compared to the experimental one. Finally, 2° AoA case (\cref{COM_Cp2}) indicates a notable disagreement both at the leading and trailing edge, which can be explained with the measurement uncertainties, especially near the trailing edge on positions 16 and 18 for the lower surface and positions 13 and 15 for the upper surface. Nevertheless, at other tapping positions, errors are smaller and consistent.

Overall, there is a possibility that the trailing edge errors, which occur in all cases at the last tapping positions, can be caused by experimental sensor error. Additionally, expected values around trailing edge should be mostly around zero, which the LBM results corroborate. Furthermore, the accuracy of measuring equipment can have a significant influence on the validity of the results. Similarly, minor numerical errors can lead to significant disparities between the experimental data and numerical results.

Compared to other scientific papers published in the last two years related to fluid flow analysis around NACA0012 airfoil, this article introduces new and efficient methodology. Conventional CFD methods, such as the Finite Volume Method (FVM) and Finite Element Method (FEM), are more often used for external fluid flow problems, although in recent times, novel mesoscopic approaches are being investigated. \citet{Krenchiglova2022} provides incompressible flow analysis with different collision operators, such as BGK, single relaxation time (SRT), moment-based model, and a model with two relaxation times. Different type of LBM approaches have been validated with NACA0012 airfoil for various ranges of AoA, up to 12°. Also, the field of interest has been low Reynolds number (up to 1000), so we can conclude that the validation in our study for the medium Reynolds number is a step forward in assessing LBM capabilities and applicability.

Conventional CFD methodologies have been employed for a wide range of Reynolds numbers. Paper by \citet{Sanmiguel2022} evaluated fluid flow around NACA0012 for low Re (16 000) and 8° AoA, with $\gamma$-$Re_0$ shear stress transport (SST) turbulence model. Likewise, \citet{Chang2022} analyzed low Re (10 000) case using $k$-$\epsilon$ turbulence model. The authors put an emphasis on the vortex shedding for NACA0012 airfoil at 0° to 12° AoA. With the classic $k$-$\epsilon$ turbulence model and Reynolds Stress Model (RSM), a medium Re (360 000) case has been assessed by \citet{Almohammadi2022} for up to 20° AoA. For high Re, \citet{Nived2022} investigated fluid flow with Spalart–Allmaras (SA) model, Menter’s $k$-$\omega$ shear stress transport (SST) model, $k$-$kL$ model and SA-Bas Cakmakcioglu modified (SA-BCM) transition model (\citet{Cakmakcoglu2020}).

\begin{figure}[H]
      \centering
      \includegraphics[width=0.8\columnwidth]{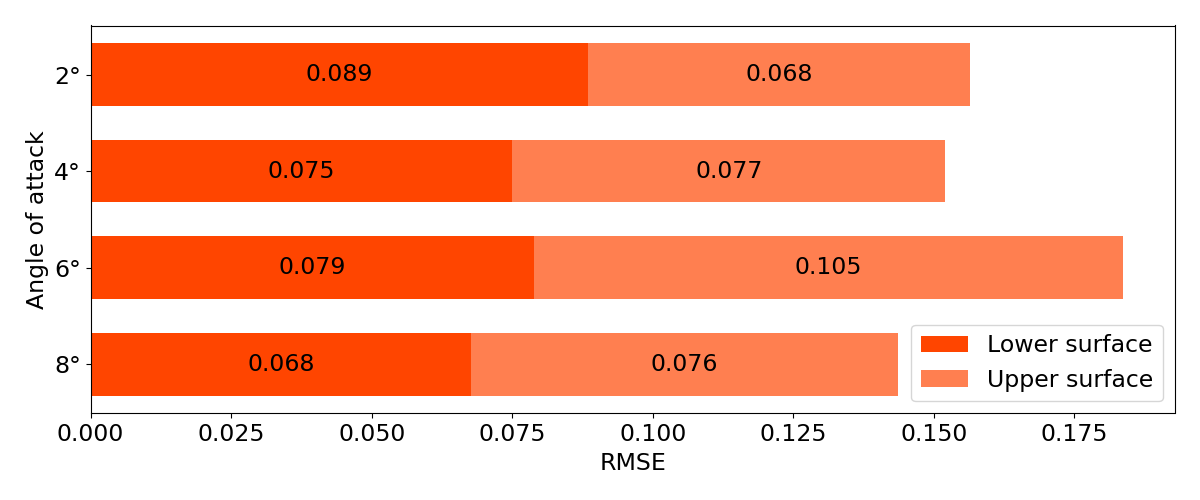}
      \caption{RMSE analsys for different AoA cases with separately lower and upper surface visualization.\label{bar}}
\end{figure}

This research addresses a novel mesoscopic approach for medium Reynolds number problem. The results are validated with the experimental data and compared to results from the literature. Overall, RMSE between experimental and LBM results for different AoA cases according to \cref{RMSE}, in \cref{mesh_ass} suggests acceptable agreement. The aforementioned comparison is visualized in \cref{bar} by observing separately lower and upper surface of NACA0012 airfoil RMSE.
Furthermore, lower surface has smaller RMSE for all AoA cases, except for 2° AoA. Also, notable deviation on leading edge contributes to this difference. \cref{bar} indicates that better congruence is achieved as the AoA increases, with the exception of 6° AoA. Main reason are larger deviations at the trailing edge. Overall, RMSE suggests that the experimental data and LBM results are in acceptable agreement.

All reported simulations were conducted using Graphics processing unit (GPU) LBM implementation UltraFluidX. Simulations are executed on a Quadro M6000 powered system. Usage on GPU’s massively parallel architecture enables short computational times. Specifically, computational time for the finest mesh with 8° AoA is 23h 34min and 17s.

\section{Conclusion}
The goal of this study is to investigate the applicability of LBM-LES implementation utilising cumulant collision operator in combination with D3Q27 velocity set, to solve medium Reynolds number flow problem. Numerical results presented in this paper are additionally validated with experimental measurements. Experiments have been conducted in an open air wind tunnel and considered a NACA0012 airfoil under various angles of attack.

Overall, obtained results in mesh independence study correlated very well. $C_p$ RMSE with regards to the experimental data, for chosen numerical mesh and for varying angles of attack, are typically below 0.1 for upper surface and 0.09 for lower surface. Additionally, if we neglect divergence on the trailing edge, for 6° AoA, overall RMSE for upper surface is below 0.08. Therefore, we can conclude that data correlation is affirmed. Furthermore, utilised generalized wall model can, as demonstrated, reliably model the flow near the wall. Collisions are adequately resolved due to the implemented cumulant collision operator.

One of the goals of this study is to highlight the simplicity of the model setup when using LBM, as well as emphasize computational savings which, consequently, leads to quick turnaround times, in contrast to conventional CFD methods. Likewise, LBM required less preprocessing time when generating lattices, which is not the case when creating a quality FVM (or FEM) mesh. Another big advantage is GPU-associated scalability, which provides more computational power per unit and is thus more efficient compared to classical CPU-only CFD codes.

Although LBM is memory-intensive, it is highly scalable, hence with continuous computational advancements, especially with regards to massively parallel architectures, it offers an accurate and efficient alternative for external aerodynamics problems. However, LBM has also limits and disadvantages. First of all, it is not particularly efficient for simulating steady flows, due to being inherently time-dependent. Secondly, LBM is not appropriate for simulating strongly compressible flows, such as transonic and supersonic flows with higher Mach number (\citet{Succi2001, Kruger2017}). Also, it is still challenging to use LBM for various types of medium and high Re problems. Nevertheless, it is a relatively novel method under continuous development, so further advancements and applicability for general problems can be assumed.

\section{Acknowledgements}
The authors also acknowledge the support of the Center of Advanced Computing and Modelling (CNRM), the University of Rijeka, for providing supercomputing resources for numerical simulations. Likewise, the work of doctoral student Andro Rak has been fully funded by the Croatian Science Foundation. Also, this research was partially supported by the KLIMOD project funded by the Ministry of Environment and Energy of the Republic of Croatia and the European structural and investment funds under grant no. KK.05.1.1.02.0017.

\bibliographystyle{elsarticle-harv} 
\bibliography{cas-refs}

\end{document}